\documentclass[12pt]{article}

\usepackage[english]{babel}

\usepackage[letterpaper,top=2cm,bottom=2cm,left=1.5cm,right=1.5cm,marginparwidth=1.75cm]{geometry}

\usepackage{amsmath,amssymb,amsmath,mathtools,eepic}
\usepackage{xcolor,cite}
\usepackage{graphicx,psfrag}
\usepackage[colorlinks=true, linkcolor=blue, citecolor=red]{hyperref}

\usepackage{tikz-cd}
\usepackage{subcaption}

\usepackage{csquotes}

\renewcommand{\bar}{\overline}
\DeclareMathOperator{\Tr}{Tr}
\DeclareMathOperator{\tr}{tr}
\DeclareMathOperator{\Lie}{Lie}
\DeclareMathOperator{\Ad}{Ad}

\newcommand{\R}{\mathcal{R}}
\newcommand{\G}{\mathrm{G}}
\renewcommand{\H}{\mathrm{H}}
\DeclareMathOperator{\End}{End}

\numberwithin{equation}{section}

\begin{document}
\title{On dual regime in Yang-Baxter deformed $\mathrm{O}(2N)$ sigma models}
\date{}
\author{Alexey Bychkov$^{1,2}$ and  Alexey Litvinov$^{1}$
\\[\medskipamount]
\parbox[t]{0.85\textwidth}{\normalsize\it\centerline{1. Skolkovo Institute of Science and Technology, 121205 Moscow, Russia}}
\\
\parbox[t]{0.85\textwidth}{\normalsize\it\centerline{2. HSE University, 6 Usacheva str., Moscow 119048, Russia}}}
\maketitle

\begin{abstract}
In this paper, we explore a new class of integrable sigma models, which we refer to as the "dual regime" of Yang-Baxter (YB) deformed $\mathrm{O}(2N)$ sigma models. This dual regime manifests itself in the conformal perturbation approach. Namely, it is well known that conventional YB-deformed $\mathrm{O}(N)$ sigma models are described in the UV by a collection of free bosonic fields perturbed by some relevant operators. The holomorphic parts of these operators play the role of screening operators which define certain integrable systems in the free theory. All of these integrable systems depend on a continuous parameter $b$, which parametrizes the central charge, and are known to possess the duality under $b^2\longleftrightarrow -1-b^2$. Although $\mathrm{O}(2N+1)$ integrable systems are self-dual, $\mathrm{O}(2N)$ systems are not. In particular, the $\mathrm{O}(2N)$ integrable systems provide new perturbations of the sigma model type. We identify the corresponding one-loop metric and $B-$field and show that they solve the generalized Ricci flow equation.
\end{abstract}
\section{Introduction}
Nonlinear sigma models \cite{Gell-Mann:1960mvl} are field theories that are of utmost importance in modern mathematical and theoretical physics: from condensed matter physics to string theory. In two space-times dimensions, sigma models are known to be renormalizable in one loop \cite{Friedan:1980jm}. Certain examples of sigma models demonstrate similarities with realistic $4D$ gauge theories, such as  asymptotic freedom, instantons, and confinement.  

There is a large class of integrable sigma models that can potentially be solved exactly at both the classical and quantum levels. One of the most well-known examples of classical integrable sigma models is the principal chiral model (PCM), where fundamental fields are valued in the Lie group $\G$. The integrability of PCM model has been established in \cite{Zakharov:1973pp}. Another important class is the symmetric space sigma models (SSSM) with fundamental field taking values in the symmetric space coset $\G/\H$, whose classical integrability has been shown in \cite{Eichenherr:1979ci}. However, these two classes of sigma models (PCM and SSSM) do not exhaust all possible integrable sigma models. In particular, integrable deformations of known integrable sigma models have attracted a lot of attention and interest in recent years. The first example of integrable deformations studied in the literature were the trigonometric and elliptic deformations of the $\mathrm{SU}(2)$ PCM discovered in \cite{Cherednik:1981df}. Later, integrable deformations of $\mathrm{SU}(2)/\mathrm{U}(1)$ SSSM \cite{Fateev:1992tk} and $\mathrm{SU}(2)$ PCM models \cite{Fateev:1995ht,Fateev:1996ea} have been discovered,  although their classical integrability was not clear at that time and has been shown later in \cite{Lukyanov:2012zt}. 

An important class of integrable deformations, called Yang-Baxter (YB) deformations, that generalizes the models discovered in \cite{Fateev:1992tk,Fateev:1995ht,Fateev:1996ea},  has been constructed in \cite{Klimcik:2008eq,Delduc:2013fga,Klimcik:2014bta}. More general deformations were also discovered soon after. We refer an interested reader to the recent review \cite{Hoare:2021dix}.

In these notes, we will be focusing primarily on YB-SSSM $\G/\H$ theories (they are also known as $\eta$ deformed symmetric space sigma models, where $\eta$ is the deformation parameter). Being integrable at the classical level, they are also remarkable as quantum field theories. In particular,
\begin{itemize}
    \item They are  renormalizable with one running coupling $\eta$.
    \item They are asymptotically free.
    \item They are (believed to be) integrable at the quantum level, provided that $H$ is simple\footnote{For $N=1$ supersymmetric versions of these models this condition can be dropped.}
\end{itemize}
Quantum integrability is checked by considering perturbative scattering of lightest particles in the weakly coupled regime and comparing these perturbative amplitudes with exact deformed (trigonometric) $S-$matrices, which solve the quantum Yang-Baxter equation \cite{Fateev:1995ht,Fateev:1996ea,Fateev:2018yos,Fateev:2019xuq}. Assuming quantum integrability to hold in all loops, one can argue that UV region of such theories can be described in conformal perturbation approach as a theory of a certain conformal field theory perturbed by a set of relevant operators \footnote{In the simplest case of deformed $\mathrm{O}(N)$ sigma-model the corresponding CFT is just the collection of free fields. For other $G/H$ sigma models, the situation might be different, and thus the corresponding CFT might be interacting.}. The holomorphic parts of these operators are known as screening operators, and they turn out to form root systems of affine Kac-Moody algebras, giving rise to Toda field theories, which are known to possess an infinite tower of mutually commuting integrals of motion. This observation leads to the following conjecture: every YB-SSSM should have an alternate description in terms of Toda field theory \cite{Fateev:2018yos,Litvinov:2018bou,Fateev:2019xuq,Alfimov:2020jpy}. 

Being restricted to the class of theories whose UV limits coincide with a collection of free fields $\varphi_j$'s, one could attempt to verify an integrability/renomalizability conjecture in at least two ways. The first is to find an appropriate UV-expansion of the metric and the $B-$field so that the screening system manifests itself: 
\begin{center}
    deformed sigma model\quad $\xlongrightarrow{\text{UV expansion}}$\quad screening system
\end{center}
\begin{enumerate}
    \item Pick a YB-SSSM, compute the metric and the $B-$field.
    \item Solve the $1$-loop RG equations (generalized Ricci flow), observe that $\eta$ is the only running coupling satisfying $\dot\eta \sim(1+\eta^2)$.
    \item Expand the metric and the $B-$field in the vicinity of the UV fixed point $\eta = i$ in powers of $(\eta-i)$ and find the corresponding relevant operators. Often they take the form $\mathcal O_{k}=A_{ij}\partial_+\varphi_i\partial_-\varphi_j e^{(\boldsymbol{\beta}_{k},\boldsymbol{\varphi})}$ with \underline{real} constant vectors $\boldsymbol{\beta}_{k}$. Imaginary $\boldsymbol{\beta}$'s give rise to irrelevant perturbations.
    \item Out of obtained relevant operators $\mathcal{O}_{k}$, reconstruct the full integrable system of screening charges. In general, it consists of different types of screening fields, either fermionic $e^{(\boldsymbol{\alpha},\boldsymbol{\varphi})}$ with $(\boldsymbol{\alpha},\boldsymbol{\alpha})=-1$, or bosonic (dressed or sigma-model type) including both real and imaginary exponential operators. This integrable system will depend on a continuous parameter $b$, such that $b^{-2}$ plays the role of the Planck constant. An example of such an integrable system is a bowtie kind diagram (see \eqref{O(2N)-bantik}, \eqref{O(2N+1)-bantik}). 
\end{enumerate}
The opposite way is to solve the generalized Ricci flow equations:
\begin{center}
    screening system \quad $\xlongrightarrow{\text{Ricci flow}}$\quad deformed sigma model
\end{center}
\begin{enumerate}
    \item Pick a screening system of a sigma-model type (for $\mathrm{O}(N)$ sigma-models it will be of bowtie type). It depends on a continuous parameter $b$.
    \item Write down the relevant perturbations in the limit $b\rightarrow 0$. These operators correspond to ''real'' exponents in $\mathcal O_{k} \sim e^{(\boldsymbol{\beta}_{k},\boldsymbol{\varphi})}$. Neglect the ''imaginary'' ones.
    \item Interpret this Toda field theory as a UV region of some sigma model and solve generalized Ricci flow equations $-\dot G_{\mu\nu} = R_{\mu\nu} + 2\nabla_\mu\nabla_\nu\Phi$ by iterations in conformal perturbation parameter. The exact solutions for the metric, $B-$field, and dilaton field are thus found.
    \item Perform T-dualities and change the target-space coordinates as needed. Observe that this sigma model coincides with a YB-SSSM\footnote{We note that the steps 3 and 4 can be united, and one can solve generalized Ricci flow equation right away \eqref{1-loop-equations}. But technically, it is much easier to follow the steps 3 and 4 instead.}.
\end{enumerate}

A remarkable property of bowtie type integrable systems responsible for YB-deformed $\mathrm{O}(N)$ sigma-models is the duality $b^2 \longleftrightarrow -1-b^2$, which holds at the level of Toda field theory \cite{Fateev:2018yos, Fateev:2019xuq,Litvinov:2018bou}. This duality swaps ''real'' and ''imaginary'' operators in the full screening system, and thus it gives rise to some duality on the sigma model side:
\begin{equation}\label{main-diagram}
    \begin{tikzcd}[column sep=5cm]
    \boxed{\text{YB-SSSM}} \arrow[d, dashed, "\text{DUAL REGIME}"'] \arrow[r, "\text{UV expansion}"] & \boxed{\text{set of relevant operators}}\arrow[d, "\text{swap } b^2\longleftrightarrow-1-b^2"] \\
    \boxed{\text{dual regime of YB-SSSM}}  & \arrow[l, "\text{Ricci flow}"] \boxed{\text{dual set of relevant operators}}
\end{tikzcd}
\end{equation}
The leftmost vertical arrow is the arrow we are interested in. Following it, we will find new solutions to generalized Ricci flow equations and show that they can be quite simply obtained from the ordinary YB-deformed $\mathrm{O}(N)$ sigma models by setting certain coordinates to constant values.

The plan of this paper is the following. In section \ref{Preliminaries} we review known results about YB-deformed $\mathrm{O}(N)$ sigma models, including explicit solution of Ricci flow equation and the relation to bowtie integrable systems of screening operators. In Section \ref{Main} we formulate our main result: a solution of the generalized Ricci flow equation for $\mathrm{O}(2N+1)$ sigma models in the dual regime (see \eqref{main-diagram}). Then in section \ref{concl} we provide some concluding remarks and highlight the direction for future work. 
\section{Preliminaries}\label{Preliminaries}
Sigma models are classically described by smooth maps $X:\Sigma\longrightarrow M$, where $\Sigma$ is a Riemannian manifold called the world-sheet and $M$ is a smooth manifold called the target-space. It is assumed to be equipped with a (preudo-)Riemannian metric $G$ and possibly other geometric data such as Kalb-Ramond 2-form field $B$, dilaton, higher-form fields consistent with the metric, etc. The action of such a theory can be written in the Polyakov form \cite{POLYAKOV1981207}
\begin{equation}
	S = \frac{1}{4\pi}\int \Big( \sqrt{|h|}h^{\alpha\beta}\, G_{\mu\nu}(X)\, \partial_\alpha X^\mu\, \partial_\beta X^\nu + \epsilon^{\alpha\beta} B_{\mu\nu}(X)\,\partial_\alpha X^\mu\, \partial_\beta X^\nu + \sqrt{|h|}R^{(2)}\,\Phi(X)+\dots \Big)\,d^2x,
\end{equation}
where $h_{\alpha\beta}$ is a metric on a two-dimensional world-sheet $\Sigma$, $G_{\mu\nu}$ and $B_{\mu\nu}$ are metric and $B$-field, $R^{(2)}$ is Ricci curvature of $h$ and $\Phi$ is a dilaton field. All these geometric data on $M$ are pulled back on $\Sigma$ via the field map $X:\Sigma\longrightarrow M$, written in local coordinates as $X^\mu(x)$, and then integrated over $\Sigma$ against the usual Lebesgue measure. In this work, we will just take $\Sigma$ to be a Minkowski flat space (a real 2-plane or a cylinder) with the light-cone coordinates $x^\pm = x^0 \pm x^1$. With this choice the sigma model action involving the metric and the $B$-field takes the following form ($\partial_\pm = \frac{\partial_0 \mp \partial_1}{2}$):
\begin{equation}
	S = \frac{1}{4\pi}\int \Big(G_{\mu\nu}(X) + B_{\mu\nu}(X)\Big)\,\partial_+ X^\mu\, \partial_- X^\nu \ d^2x.
\end{equation}

Two-dimensional non-linear sigma models are renormalizable. At $1$-loop order the RG equations \cite{Friedan:1980jm} (see also \cite{Braaten:1985is}) are often called the generalized Ricci flow equations and read
\begin{equation}\label{1-loop-equations}
\begin{aligned}
	-\dot{G}_{\mu\nu} &= R_{\mu\nu}-\frac{1}{4}H_{\mu}^{\kappa\lambda}H_{\nu\kappa\lambda}+\nabla_{\mu}V_{\nu}+\nabla_{\nu}V_{\mu},\\
	-\dot B_{\mu\nu} &= H_{\lambda\mu\nu}V^{\lambda}-\frac{1}{2}\nabla_{\lambda}H^{\lambda}_{\ \mu\nu}+\nabla_{\mu}\omega_{\nu}-\nabla_{\nu}\omega_{\mu},
\end{aligned}
\end{equation}
where $V_\mu$ and $\omega_\mu$ correspond to a wave function renormalization. By $\nabla_\mu$ we denote the covariant derivative with respect to the usual metric connection; $H = dB$ is the magnetic flux $3$-form.
\subsection{PCM and sigma models on symmetric spaces}
\subsubsection*{PCM is integrable and renormalizable}
We start by reviewing general classical and quantum properties of nonlinear sigma models on arguably the simplest non-trivial example, the so-called \textit{principal chiral model}. The principal chiral model (PCM) is the non-linear sigma model with the target space being $M = \G$, a compact semisimple Lie group equipped with the canonical metric $ds^2 = -\lambda^{-2}\tr\big(g^{-1}dg\,g^{-1}dg\big)$. Here
$\lambda^2$ is some positive constant (overall scale) and $\tr$ denotes the Cartan-Killing metric on $\mathfrak{g} = \Lie(\G)$, which for compact semisimple groups is strictly negative definite and thus nondegenerate. The action of PCM model takes the form
\begin{equation}
	S = -\frac{1}{4\lambda^2}\int \tr\big(g^{-1}\partial g\,g^{-1}\bar \partial g\big) d^2x.
\end{equation}

PCM is a fairly simple NLSM, yet it enjoys a plethora of nice properties. One of these is the celebrated Lax integrability \cite{Zakharov:1973pp}. It can be seen as follows. Classical equations of motion are\footnote{These are just conservation law for Noether current corresponding to right symmetry $g(z, \bar z) \mapsto g(z, \bar z)\cdot h$. This is not a coincidence.}
\begin{equation}
	\partial_+ j_- + \partial_- j_+ = 0, \qquad j_\pm = g^{-1}\partial_\pm g.
\end{equation}
Oftentimes it is more convenient to work in the coordinate-independent setting. Namely, denote $j = g^{-1}dg \in \Lambda^1(\Sigma)\otimes\mathfrak g$, so that $j_\pm$ are just the light-cone components of the current 1-form $j$. The conservation equations are then rewritten as $\partial_\alpha j^\alpha = 0$, or, in the coordinate-independent fashion, as $d\star j = 0$. Now, since $j\in\Lambda^1(\Sigma)\otimes\mathfrak{g}$ is a pull-back of the canonical 1-form  on the group via the section $g:\Sigma\longrightarrow \G$, it is also flat: $dj + \frac{1}{2}[j\wedge j] = 0$.

Consider the following Lie algebra-valued $1$-form:
\begin{equation}\label{PCM-Lax-connection}
	L(z) = \frac{j + z\star j}{1+z^2} \qquad \text{or}\qquad L_\pm(z) = \frac{j_\pm}{1\mp z},\qquad z\in\mathbb{C}.
\end{equation}
It depends on the spectral parameter $z$ and is flat if and only if the equations of motion are satisfied:
\begin{equation}
	dL(z) + \frac{1}{2}\left[L(z)\wedge L(z)\right] = 0 \qquad \Longleftrightarrow\qquad
	\begin{cases}
		d\star j = 0\\
		dj + \frac{1}{2}[j\wedge j] = 0
	\end{cases}
\end{equation}
We call a mechanical system Lax integrable precisely when it admits the so-called Lax connection $L(z)$, flatness of which is equivalent to the classical equations of motion.

The $1$-loop renormalization group flow in sigma models without Kalb-Ramond term and dilaton is given by the celebrated Ricci flow equation \cite{Friedan:1980jm}:
\begin{equation}
	\frac{d G_{\mu\nu}}{dt} = - R_{\mu\nu}, \qquad t = \log\frac{\Lambda_0}{\Lambda},
\end{equation}
where UV region is $t\rightarrow -\infty$ and IR region is $t\rightarrow 0$. It can be shown that the canonical metric on the Lie group is in fact Einstein metric:
\begin{equation}
	ds^2 = -\frac{1}{\lambda^2} \Tr\big(g^{-1}dg\,g^{-1}dg\big) \quad\Rightarrow\quad R_{\mu\nu} = h^\vee(G)\cdot G_{\mu\nu},
\end{equation}
where $h^\vee(\G)$ is the dual Coxeter number of Lie group $\G$. We conclude that in the principal chiral model the only running coupling constant is $\lambda$:
\begin{equation}
	\frac{d}{dt}\left(\frac{1}{\lambda^2}\right) = h^\vee(G) \quad \Rightarrow \quad \frac{1}{\lambda^2(t)} = -h^\vee(G)\cdot t.
\end{equation}
So, in the UV limit $t\longrightarrow -\infty$ the theory becomes free: $\lambda(-\infty) = 0$, and in the IR region $t\rightarrow0$ it becomes strongly coupled. This phenomenon is widely known as asymptotic freedom.
\subsubsection*{Symmetric $\G / \H$ models are integrable and renormalizable}
It is possible to construct a sigma model on homogeneous space $\G/\H$ by gauging out $\H$-degrees of freedom. For this we introduce $\mathfrak{h} = \Lie(\H)$-valued gauge field
\begin{equation}
	\partial_\mu \longmapsto \partial_\mu + A_\mu,\quad A\in\Lambda^1(\Sigma)\otimes\mathfrak{h},
\end{equation}
plug it into the PCM action, and then integrate $A$ out (which is a purely classical procedure, since $A$ does not have a kinetic term). Since $\mathfrak g$ is compact (semi-)simple, it can be decomposed into the orthogonal direct sum $\mathfrak g = \mathfrak h \overset{\perp}{\oplus} \mathfrak m$. The result of integration over $A_\pm\in\mathfrak h$ is
\begin{equation}\label{G/H-non-deformed}
	S_{\G/\H} = -\frac{1}{4\lambda^2}\int \tr\Big((g^{-1}\partial_+ g)^{\mathfrak m}\ (g^{-1}\partial_- g)^{\mathfrak m}\Big)d^2x = -\frac{1}{4\lambda^2}\int\tr\Big(g^{-1}\partial_+ g\ \mathrm{P}_\mathfrak m(g^{-1}\partial_- g)\Big)d^2x,
\end{equation}
where $\mathrm P_\mathfrak{m}$ is a orthogonal projector on $\mathfrak m = \mathfrak{h}^\perp$, and $j^{\mathfrak m} = \mathrm P_{\mathfrak m}(j)$ is $\mathfrak{m}$-component of $j$.

The metric on the target space $\G/\H$ corresponding to the action
\eqref{G/H-non-deformed} is obviously $\G$-invariant. Consider its restriction on the tangent space over a single point $p_0 = e\cdot\H\in \G/\H$. At this point the metric in question is (up to multiplication by $\lambda^{-2}$) just the usual Cartan-Killing form on $\mathfrak g = \mathrm{T}_e\G$ restricted onto $\mathfrak h^\perp = \mathfrak m$. The transitive action of $\G$ on $\G/\H$ can be used to construct a global metric on the homogeneous space.

The property of classical (Lax) integrability of PCM remains intact after gauging for some homogeneous spaces. More precisely, if the Lie algebra $\mathfrak{g} = \mathfrak{h} \oplus \mathfrak{m}$ admits an automorphism $\sigma$ of order $2$, which preserves $\mathfrak h$ subalgebra
\begin{equation}
	\sigma(\mathfrak h) = \mathfrak h; \qquad \sigma(\mathfrak m) = -\mathfrak m,
\end{equation}
then the coset $\G/\H$ is called symmetric homogeneous space, and the corresponding homogeneous space sigma model is Lax integrable with the Lax connection
\begin{equation}
	L(z) = A + \frac{1-z^2}{1+z^2}K + \frac{2z}{1+z^2}\star K, \qquad \text{where}\quad  A = \mathrm{P}_\mathfrak{h}(j),\quad K = \mathrm{P}_\mathfrak{m}(j).
\end{equation}

RG flows in the symmetric space sigma models are of no difference from those in the principal chiral models, since every symmetric homogeneous space is in fact Einstein too!
\subsection{Yang-Baxter deformation}
We recall that the principal homogeneous space $\G$ endowed with the canonical metric is Einstein (meaning that the Ricci tensor is proportional to the metric). $\G/\H$ is Einstein provided $\G/\H$ is symmetric. Since 1-loop RG equation reads $-\dot G_{\mu\nu} = R_{\mu\nu}$, the target-space metric $G_{\mu\nu}$ depends on RG time $t$ only through the overall scale $\lambda^{-2}(t)$,
\begin{equation}
	G_{\mu\nu}(t) = \lambda^{-2}(t)\cdot G_{\mu\nu}(t_0),
\end{equation}
so, there is in fact only one running coupling constant in these theories. We remind the reader that the general sigma model is a theory with an infinite number of couplings, since $G_{\mu\nu}(X)$ is a function of target-space coordinates, which can be as complicated as one wants it to be.

A more interesting dependence on $t$ can be obtained by introducing another independent coupling. A very fruitful way to do so is to consider integrable parametric deformations. "To deform a theory" means making the metric and $B$-field depend on a finite number of additional deformation parameters (in this work we will only add one), which are then identified with new coupling constants. Sometimes, it is possible to make the deformation preserve classical integrability, in which case the deformation is called integrable. 

One of the main mysteries about these integrable deformations is the integrability-renormalizability conjecture \cite{Fateev:1992tk,Fateev:1995ht,Fateev:1996ea,Lukyanov:2012zt}. Assume that the metric and the $B$-field in certain sigma model depend only on a finite number of parameters $\{\lambda_1, \lambda_2, ..., \lambda_N\}$:
\begin{equation}
	G_{\mu\nu} = G_{\mu\nu}\big(\{\lambda_k\}\big), \qquad B_{\mu\nu} = B_{\mu\nu}\big(\{\lambda_k\}\big),
\end{equation}
and that the sigma model with these $G_{\mu\nu}$ and $B_{\mu\nu}$ is classically integrable. Then, under renormalization $G_{\mu\nu}$ and $B_{\mu\nu}$ will depend on RG-time only through these couplings $\lambda_k(t)$'s. We call such sigma models \textit{RG-stable} or renormalizable in the strong sense. In short, this conjecture states the following:
\[
\text{RG-stability} \qquad \Longleftrightarrow\qquad \text{Integrability}
\]
\subsubsection*{Yang-Baxter PCM}
One of these integrable deformations is the so-called Yang-Baxter deformation of PCM \cite{Klimcik:2008eq}. In what follows, we will describe the construction and the basic features of it. We start from the action
\begin{equation}\label{YB-PCM-action}
	S = -\frac{1}{4\lambda^2}\int \tr\left(j_+ \frac{1}{1 - \eta \mathcal{R}_g}j_-\right)d^2x,\qquad \R_g = \Ad_{g}^{-1}\R\Ad_g,
\end{equation}
where $\R\in\End(\mathfrak g)$ is a constant linear operator on the Lie algebra, $\eta\in\mathbb C$ is the deformation parameter, and the limit $\eta \rightarrow 0$ corresponds to the non-deformed model.
\subsubsection*{Equations of motion.} Equations of motion for YB-PCM can be obtained by varying $g(x) \mapsto g(x)(1-\varepsilon(x))$ (infinitesimal right-acting symmetry). It is more convenient to rewrite the Lagrangian in the following way: since $\R_g = \Ad_g^{-1}\R\Ad_g$, 
\begin{multline}
	L = \tr\left(j_+\frac{1}{1-\eta\R_g}j_-\right) = \tr\left(j_+\cdot \Ad_g\frac{1}{1-\eta\R}\Ad_g^{-1}j_-\right) =\\
	=\tr\left(\Ad_g^{-1}(j_+)\cdot\frac{1}{1-\eta\R}\Ad_g^{-1}(j_-)\right) = \tr\left(k_+\frac{1}{1-\eta\R}k_-\right),
\end{multline}
where $k_\pm$ are the components of the right-invariant current:
\begin{equation}
	k = -dg\,g^{-1}; \qquad \Ad_g^{-1}(j) = - k.
\end{equation}
Under the infinitesimal right shifts $\delta g = -g\varepsilon$ these currents transform as $\delta k = \Ad_g(d\varepsilon)$, so, by direct computation the variation of the Lagrangian is
\begin{equation}
	\delta L = \tr\Big(\varepsilon\cdot\left(\partial_+ J_- + \partial_- J_+\right)\Big) + \text{total derivatives},
\end{equation}
where
\begin{equation}
	J_- = \frac{1}{\xi}\Ad_g^{-1}\frac{1}{1-\eta\R}k_- =\frac{1}{\xi} \frac{1}{1-\eta\R_g}j_-,\qquad
	J_+ = \frac{1}{\xi}\Ad_g^{-1}\frac{1}{1-\eta\R^\mathrm{t}}k_+ = \frac{1}{\xi}\frac{1}{1-\eta\R^\mathrm{t}_g}j_+.
\end{equation}
Here, $\R^\mathrm{t}\in\End\mathfrak{g}$ is the operator conjugated to $\R$ with respect to the Killing form: $\tr(X\, \R(Y)) = \tr(\R^\mathrm{t}(X)\,Y)$, and $\xi$ is an arbitrary number.
\subsubsection*{Classical integrability.} If we require $J_\pm$ to be not only conserved, but also flat
\begin{equation}\label{YB-PCM-flatness-cond}
	\partial_+ J_- - \partial_- J_+ + [J_-, J_+] = 0,
\end{equation}
then the Lax connection could be defined by $L_\pm(\lambda) = \frac{J_\pm}{1\mp\lambda}$ (exactly as in (\ref{PCM-Lax-connection})) and the system would be Lax integrable. The direct computation shows that the flatness condition (\ref{YB-PCM-flatness-cond}) is equivalent to the system of two equations on $\R\in\End\mathfrak{g}$:
\begin{equation}\label{mcYBE}
	\begin{cases}
		\R + \R^\mathrm{t} = 0;\\
		\left[\R X, \R Y\right] + \R\big([X, \R Y] + [\R X, Y]\big) + c^2[X, Y] = 0, \quad \forall X, Y\in\mathfrak{g},
	\end{cases}
\end{equation}
where $c^2$ is an arbitrary complex parameter. The first equation forces $\R$ to be skew-symmetric. The second equation is called \textit{modified classical Yang-Baxter equation} (mcYBe). Usual (unmodified) cYBe is recovered at $c^2 = 0$. For general $c$ we can always rescale $\R$ by a real factor to make $c^2\in\mathbb R$ equal to $1$ or $-1$ or $0$. The solutions to these mcYBe's are usually called unmodified ($c = 0$), non-split $(c = \sqrt{-1})$ and split $(c = 1)$ $r$-matrices.
\subsubsection*{Drinfeld-Jimbo solution} One solution to mcYBe is the so-called Drinfeld-Jimbo $r$-matrix $\R^\mathrm{DJ}$. Choose a Cartan-Weyl basis in $\mathfrak{g} = \langle E_\alpha, E_{-\alpha}, H_i\rangle$. In this basis one has
\begin{equation}
	\R^\mathrm{DJ}(E_{\pm\alpha}) = \pm c\, E_{\pm\alpha}, \qquad \R^\mathrm{DJ}(H_i) = 0.
\end{equation}
We notice that in the non-split case $\R^3 = -\R$, which gives rise to a non-zero $B$-field. Indeed, expanding the geometric series
\begin{equation}
	\frac{1}{1-\eta\R} = \sum_{k=0}^{+\infty} \eta^{2k}\R^{2} + \sum_{k=0}^{+\infty} (-1)^k\eta^{2k+1}\R = 1 + \frac{\eta^2}{1+\eta^2}\R^2 + \frac{\eta}{1+\eta^2}\R,
\end{equation}
we see that there is a non-zero skew-symmetric part. 
\subsubsection*{1-loop RG flow}	As it was mentioned previously, classical integrability makes RG flows in YB-PCM act only on the coupling constants $\lambda$ and $\eta$. 1-loop equations \eqref{1-loop-equations} can be shown to be \cite{Squellari:2014jfa,Sfetsos:2015nya}
\begin{equation}\label{YB-PCM-RG-flow}
	\frac{d\eta}{dt} = \mathrm{const}\,(1+\eta^2)^2, \qquad \frac{d}{dt}\left(\eta\cdot\lambda^2\right) = 0.
\end{equation}
So, there is effectively only one running coupling $\eta$, since $\lambda^2\cdot\eta$ is a first integral of RG flow.
\subsubsection*{Yang-Baxter $\G/\H$ models are integrable and renormalizable}
Since the global group of YB-PCM  contains $\G$-subgroup acting from the right, we can still gauge some $\H\subset\G$ to obtain a $1$-parametric deformation of the sigma model on the homogeneous space $\G/\H$. Replacing $d\longmapsto d + A$ in with $A\in\Lambda^1\Sigma\otimes\mathfrak{h}$ in the YB-PCM action, we get
\begin{equation}\label{YB-G/H-with-A}
	S = -\frac{1}{4\lambda^2}\int \tr\left(K_+\frac{1}{1-\eta\R}K_-\right)d^2x, \qquad K_\pm = -g\left(\partial_\pm + A_\pm\right)g^{-1}.
\end{equation}
Then we integrate $A$ out, after a simple (yet somewhat lengthy computation) we obtain the action of the Yang-Baxter $\G/\H$-model:
\begin{equation}
	S = -\frac{1}{4\lambda^2}\int \tr\left((j_+)^\mathfrak{m} \frac{1}{1-\eta\R_g\,\mathrm P_\mathfrak{m}}(j_-)^\mathfrak{m}\right)d^2x, \qquad (j_\pm)^\mathfrak{m} = \mathrm{P}_\mathfrak{m}(j_\pm),
\end{equation}
where $\R_g$ is the same as in \eqref{YB-PCM-action}. Sometimes it is more convenient to replace $\eta = i\kappa$ and $\lambda^2\kappa =: \hbar$, so that $\hbar$ doesn't run and the YB-deformed $\G/\H$-sigma model action becomes
\begin{equation}\label{YB-G/H-kappa-action}
S = -\frac{\kappa}{\hbar}\int\tr\left((g^{-1}\partial_+g)^\mathfrak{m}\,\frac{1}{1-i\kappa\,\R_g\,\mathrm{P}_\mathfrak{m}}\,(g^{-1}\partial_-g)^{\mathfrak{m}}\right)d^2x.
\end{equation}
\subsubsection*{Lax integrability} The usual way to derive the Lax connection is to find a current that is flat and conserved. It was done for symmetric spaces in \cite{Delduc:2013fga} (see also \cite{Bykov:2016pfu}). The result is
\begin{equation}
	L_\pm(z) = A_\pm + \sqrt{1-\kappa^2}\frac{J_\pm}{1\mp z},
\end{equation}
where $A_\pm$ is the $\mathfrak{h}$-valued connection $1$-form used to gauge out $\H$-degrees of freedom, and
\begin{equation}
	J_\pm = \frac{1}{1\pm\eta\R_g}j_\pm, \qquad j_\pm = g^{-1}\partial_\pm g.
\end{equation}
\subsubsection*{RG flows} Similarly to YB-PCM theory, in the YB-deformed symmetric space sigma model the only running coupling is $\kappa$, and RG flow boils down to an ODE
\begin{equation}\label{2-sausage-RG}
	\frac{d\kappa}{dt} = C\hbar\,(\kappa^2-1),
\end{equation}
where $C=h^{\vee}_{G}$ is the dual Coxeter number of $G$. Notice the difference compared to \eqref{YB-PCM-RG-flow}: in the \textit{r.h.s.} there is a power 1 of $(1-\kappa^2)^1$. For YB-PCM it was $(1-\kappa^2)^2$ \eqref{YB-PCM-RG-flow}. This difference will be of utmost importance for us in this work.

The $1$-loop equations \eqref{2-sausage-RG} are solved by
\begin{equation}
	\kappa(t) = -\tanh (C\hbar\, t)
\end{equation}
We will be interested in the UV limit ($t\rightarrow +\infty$) along this flow. There are three interesting limits:
\begin{enumerate}
	\item $\hbar\rightarrow 0$, $\kappa$ fixed\\
	This is the classical limit: the RG time along the flow is $t\sim\frac{1}{\hbar} \rightarrow -\infty$
	\item $\hbar\rightarrow 0$, $t$ fixed\\
	This is a non-deformed limit. In fact, along the flow $\kappa\sim C\hbar$, so in the general action of YB-$G/H$ \eqref{YB-G/H-kappa-action} a deformation parameter $\kappa$ is small, while the overall factor $\frac{\kappa}{\hbar}$ is finite. 
	\item $t\rightarrow -\infty$, $\hbar$ fixed\\
	This limit is the most interesting one. From general physical grounds, it can be argued that in the vicinity of UV fixed point ($\kappa \rightarrow 1$, $t \rightarrow -\infty$) the metric behaves as
	\begin{equation}
		ds^2 = ds^2_\text{flat or CFT} + \Lambda\sum\limits_\text{finite}\Big(\text{relevant perturbations}\Big).
	\end{equation}
	$\Lambda$ is a dynamically generated mass scale. The structure of relevant perturbations in the \textit{r.h.s.} is the subject of our interest.
\end{enumerate}
\subsection{Ricci flow, conformal perturbations and Toda field theories}
As it was mentioned in the previous section, in the vicinity of the UV fixed point $\kappa = 1$ the metric in asymptotically free sigma model can be approximated by
\begin{equation}
	ds^2\Big|_{\kappa = 1 + \varepsilon} = \delta_{\mu\nu}\, dX^\mu dX^\nu + \Lambda \sum\limits_r A_{\mu\nu}^{(r)}(X)\, dX^\mu dX^\nu + O(\Lambda^2),
\end{equation} 
The matrices $A^{(r)}_{\mu\nu}$ define IR-relevant operators. By physical arguments it is clear that these are the graviton vertex operators 
\begin{equation}
	\mathcal{O}^{(r)} = A_{\mu\nu}^{(r)}\,\partial_+ X^\mu\,\partial_- X^\nu\,e^{(\boldsymbol\beta_r, \boldsymbol X)},
\end{equation}
such that $\boldsymbol{\beta}_r^2>0$. In string theory they are interpreted exactly as metric perturbations of a conformal background.

In fact, in deformed $\mathrm{O}(N)$ models (YB-deformed $\mathrm{SO}(N+1)/\mathrm{SO}(N)$ models) by performing suitable T-dualities and coordinate changes it is always possible to force the matrices $A_{\mu\nu}^{(r)}$ to be of rank $1$. More precisely, $A_{\mu\nu}^{(r)} = a_{\mu}^{(r)}a_{\nu}^{(r)}$, where $\boldsymbol{a}^{(r)} = (a_{\mu}^{(r)})$ are constant ''light-like'' (of zero euclidean length) vectors:
\begin{equation}
    \boldsymbol{a}^{(r)} = (0, ..., 0, 1, 0, ..., 0, i, 0, ..., 0)
\end{equation}
A condition of perturbation to be relevant is, in fact, a very strong one: only very special choices of $\boldsymbol{\alpha}$'s and $(A_{\mu\nu})$'s are allowed. We will return to this discussion in the next section.
\subsubsection*{Example: sausage model} Now we demonstrate how these relevant operators appear in the simplest non-trivial example, YB-deformed $\mathrm{O}(3)$-model. The sausage metric and $B$-field are (see below)
\begin{equation}\label{s2-sausage-metric}
	ds^2 = \frac{\kappa}{\hbar}\frac{d\theta^2 + \cos^2\theta\,d\phi^2}{1-\kappa^2\sin^2\theta}, \qquad 
    B = -\frac{i\kappa^2}{2\hbar}\left[\frac{\sin2\theta_1}{1-\kappa^2\sin^2\theta_1}d\theta_1\wedge d\phi_1\right].
\end{equation}
For general $\kappa$ this metric solves the Ricci flow equation provided $V = \omega = 0$ and $\kappa = -\tanh\hbar t$ (see \ref{2-sausage-RG}). We perform T-duality \cite{Buscher:1987sk} in $\phi$ direction. Then the dual metric and $B$-field read
(we set $\hbar = 1$):
\begin{equation}
    ds^2_\text{T-dual} = \kappa\left[ d\theta^2 + 2i\tan\theta\,d\theta d\phi + \frac{1-\kappa^2\sin^2\theta}{\kappa^2\cos^2\theta}d\phi^2 \right], \qquad B_\text{T-dual} = 0.
\end{equation}
They solve 1-loop RG flow equations with\footnote{
Alternatively, one could notice that the $B$-field of \eqref{s2-sausage-metric} is a closed form ($H = dB = 0$), and thus does not run in RG flow itself ($\dot B_{ij} = 0$) and does not participate in the running of the metric on RG flow (see \eqref{1-loop-equations}). One could argue that $B$-field can be dropped in the T-dual case too. Indeed: we perform T-duality ignoring the closed $B$-field:
\begin{equation}
    ds^2_\mathrm{T-dual, B=0} = \frac{\kappa}{1-\kappa^2\sin^2\theta}d\theta^2 + \frac{1-\kappa^2\sin^2\theta}{\kappa\cos^2\theta}d\phi^2.
\end{equation}
This metric solved Ricci flow with
\begin{equation}
    V_\mathrm{T-dual, B=0} = \frac{(1-\kappa^2)\tan\theta}{1-\kappa^2\sin^2\theta} d\theta.
\end{equation}
}
\begin{equation}
    V_\text{T-dual} = \frac{i(1-\kappa^2)}{\kappa^2\cos^2\theta}d\phi, \qquad \omega_\text{T-dual} = 0, \qquad \text{and}\qquad \kappa_\text{T-dual} = \kappa = -\tanh t.
\end{equation}
We now perform the following convenient change of coordinates:
\begin{equation}
    \phi \longmapsto \phi + \frac{i}{2}\log\sinh^2 y,\qquad
    \theta \longmapsto \arcsin\coth y.
\end{equation}
The metric becomes
\begin{equation}
    ds^2_\mathrm{T-dual} = \frac{1+\kappa^2}{2\kappa}(dy^2 + d\phi^2) + \frac{1-\kappa^2}{4\kappa}\Big[ (dy + i\,d\phi)^2 e^{2y} + (dy - i\,d\phi)^2 e^{-2y} \Big].
\end{equation}
Since T-duality preserves $t$-dependence in $\kappa(t)$, we write for T-dual metric
\begin{equation}
\kappa = -\tanh t = \frac{1 - e^{2t}}{1 + e^{2t}} = 1 - \epsilon + O(\epsilon^2), \quad \text{where } \epsilon = \epsilon(t) := 2\,e^{2t},
\end{equation}
which results in the following UV-expansion:
\begin{equation}
    ds^2_\text{T-dual} = dy^2 + d\phi^2 + \frac{\epsilon}{2}\, \Big[ (dy + i\,d\phi)^2 e^{2y} + (dy - i\,d\phi)^2 e^{-2y} \Big] + O(\epsilon^2).
\end{equation}
So the Toda perturbation operators take the form
\begin{equation}
    \mathcal{O}_{1,2} = \int A_{\mu\nu}^{(1,2)}\,\partial_+ X^\mu\,\partial_- X^\nu\,e^{(\boldsymbol\alpha_{1,2}, \boldsymbol X)}\  d^2x, \qquad \boldsymbol{X} = (y, \phi)
\end{equation}
with
\begin{equation}
    A_{\mu\nu}^{(1)} = 
    \begin{pmatrix}
        1&i\\
        i&-1
    \end{pmatrix} = 
    (1,i)^{\otimes 2},
     \quad  
    \boldsymbol{\alpha}_1 = (2,0),\quad
    A_{\mu\nu}^{(2)} = 
    \begin{pmatrix}
        1&-i\\
        -i&-1
    \end{pmatrix} =(1,-i) ^{\otimes 2},
     \quad  
    \boldsymbol{\alpha}_2 =(-2,0).
\end{equation}
\subsection{Yang-Baxter \texorpdfstring{$\mathrm{O}(N)$}{O(N)} sigma models: known results}
\subsubsection*{Metric and $B$-field}
Here we follow \cite{Alfimov:2020jpy}. We use the following parametrization of the coset elements for even-dimensional spheres:
\begin{equation}
\begin{gathered}	
    \begin{aligned}
		g=e^{\phi_{1}T_{12}}e^{\theta_{1}T_{13}} &\quad\text{for}\quad \mathrm{SO}(3)/\mathrm{SO}(2),\\
		g=e^{\phi_{2}T_{34}}e^{\theta_{2}T_{35}}e^{\phi_{1}T_{12}}e^{\theta_{1}T_{13}} &\quad\text{for}\quad \mathrm{SO}(5)/\mathrm{SO}(4),\\
		g=e^{\phi_{3}T_{56}}e^{\theta_{3}T_{57}}e^{\phi_{2}T_{34}}e^{\theta_{2}T_{35}}e^{\phi_{1}T_{12}}e^{\theta_{1}T_{13}}& \quad\text{for}\quad \mathrm{SO}(7)/\mathrm{SO}(6),
	\end{aligned}\\
    \dots\dots\dots\dots\dots\dots\dots\dots\dots\dots\dots
\end{gathered}
\end{equation}
and for odd-dimensional ones:
\begin{equation}
\begin{gathered}
	\begin{aligned}
		g=e^{\phi T_{34}}e^{\phi_{1}T_{12}}e^{\theta_{1}T_{13}}& \quad\text{for}\quad \mathrm{SO}(4)/\mathrm{SO}(3),\\
		g=e^{\phi T_{56}}e^{\phi_{2}T_{34}}e^{\theta_{2}T_{35}}e^{\phi_{1}T_{12}}e^{\theta_{1}T_{13}}& \quad\text{for}\quad \mathrm{SO}(6)/\mathrm{SO}(5),\\
		g=e^{\phi T_{78}}e^{\phi_{3}T_{56}}e^{\theta_{3}T_{57}}e^{\phi_{2}T_{34}}e^{\theta_{2}T_{35}}e^{\phi_{1}T_{12}}e^{\theta_{1}T_{13}}& \quad\text{for}\quad \mathrm{SO}(8)/\mathrm{SO}(7),
	\end{aligned}
    \\
    \dots\dots\dots\dots\dots\dots\dots\dots\dots\dots\dots
\end{gathered}
\end{equation}

The YB-deformed metric and $B$-field on even-dimensional sphere $\mathbb{S}^{2N}$ have the form (here and below we set $\hbar=1$)
\begin{equation}\label{S(2N)-full-metric}
	ds^{2}_{\mathrm{YB}-\mathbb{S}^{2N}}=\sum_{k=1}^{N}S_{k}(\boldsymbol{\theta})\Bigl(d\theta_{k}^{2}+\cos^{2}\theta_{k}d\phi_{k}^{2}\Bigr),\qquad
	B_{\mathrm{YB}-\mathbb{S}^{2N}}=\sum_{k=1}^{N}b_{k}(\boldsymbol{\theta})d\theta_{k}\wedge d\phi_{k},
\end{equation}
and for odd-dimensional $\mathbb{S}^{2N+1}$:
\begin{equation}\label{S(2N+1)-YB-full-metric}
	ds^{2}_{\mathrm{YB}-\mathbb{S}^{2N+1}}=\sum_{k=1}^{N}S_{k}(\boldsymbol{\theta})\Bigl(d\theta_{k}^{2}+\cos^{2}\theta_{k}d\phi_{k}^{2}\Bigr) + S_{0}(\boldsymbol{\theta})d\chi^{2},\qquad
	B_{\mathrm{YB}-\mathbb{S}^{2N+1}}=\sum_{k=1}^{N}b_{k}(\boldsymbol{\theta})d\theta_{k}\wedge d\phi_{k}.
\end{equation}
The functions $S_{k}(\boldsymbol{\theta})$ are
\begin{equation}\label{S,S0-ans}
	S_{k}(\boldsymbol{\theta})=\kappa\frac{\prod_{j=1}^{k-1}\sin^{2}\theta_{j}}{1-\kappa^{2}\left(\prod_{j=1}^{k-1}\sin^{4}\theta_{j}\right)\sin^{2}\theta_{k}},\qquad
	S_{0}(\boldsymbol{\theta})=\kappa\prod_{k=1}^{N}\sin^{2}\theta_{k},
\end{equation}
and
\begin{equation}\label{b-ans}
	b_{k}(\boldsymbol{\theta})=-\frac{i\kappa^2}{2}\frac{\partial}{\partial\theta_{k}}\log S_{k}(\boldsymbol{\theta}) = -\frac{i\kappa^2}{2}\frac{\prod_{j=1}^{k-1}\sin^4\theta_j \sin(2\theta_k)}{1-\kappa^2\prod_{j=1}^{k-1}\sin^4\theta_j\sin^2\theta_k}d\theta_k\wedge d\phi_k.
\end{equation}
Notice how even-dimensional metric and $B$-field (\ref{S(2N)-full-metric}) are obtained from odd-dimensional ones (\ref{S(2N+1)-YB-full-metric}) by ''erasing'' the isometric $\chi$-direction.

The corresponding vector fields (or rather $1$-forms) $V = V_\mu(x) dx^\mu$ and $\omega = \omega_\mu(x) dx^\mu$ read:
\begin{align}
    V_{\mathbb{S}^{2N}} &= -2 \sum\limits_{k=1}^{N} \cot(\theta_{N-k+1})\Big((k-1)F_1(\boldsymbol{\theta}) - \sum\limits_{j=2}^{k} F_j(\boldsymbol{\theta})\Big)d\theta_k,\\
    V_{\mathbb{S}^{2N+1}} &= -\sum\limits_{k=1}^{N} \cot(\theta_{N-k+1})\Big((2k-1)F_1(\boldsymbol{\theta}) - 2 \sum\limits_{j=2}^{k} F_j(\boldsymbol{\theta}) - 1\Big)d\theta_k,
\end{align}
and
\begin{align}
    \omega_{\mathbb{S}^{2N}} &= -i\kappa\sum\limits_{k=1}^{n}\big(2N-2k+1\big)\Big(\prod\limits_{j=1}^{k-1}\sin^2\theta_j\Big)\cos^2(\theta_k)\,F_k(\boldsymbol{\theta})\,d\phi_k,\\
    \omega_{\mathbb{S}^{2N+1}} &= -i\kappa\sum\limits_{k=1}^{n}\big(2N-2k+2\big)\Big(\prod\limits_{j=1}^{k-1}\sin^2\theta_j\Big)\cos^2(\theta_k)\,F_k(\boldsymbol{\theta})\,d\phi_k,
\end{align}
where the functions $F_k(\boldsymbol{\theta})$ are just the denominators of $S_k(\boldsymbol{\theta})$:
\begin{equation}\label{Fk-ans}
    F_k(\boldsymbol{\theta}) = \left(1 - \kappa^2\ \Big(\prod\limits_{j=1}^{k-1}\sin^4\theta_j\Big)\sin^2\theta_k\right)^{-1}.
\end{equation}
With these $V$ and $\omega$ the $1$-loop equations \eqref{1-loop-equations} are solved provided that (for YB-deformed $\mathrm{O}(N)$ models)
\begin{equation}
    \kappa(t)=-\tanh(N-2)t
\end{equation}
\subsubsection*{Graviton operators in $\mathrm{O}(N)$ models}
An analysis performed in the previous section for sausage model can be generalized. In \cite{Litvinov:2018bou} it was shown that for general $\mathrm{O}(N)$ models graviton vertex operators take the form
\begin{equation}\label{dressed-screening}
	\mathcal{O}_{ij} = (\boldsymbol\alpha_i,\partial_+\boldsymbol\varphi)(\boldsymbol\alpha_i,\partial_-\boldsymbol\varphi) e^{(\boldsymbol\beta_{ij},\boldsymbol{\varphi})},
\end{equation}
where $\boldsymbol\varphi = (X^1, \dots, X^{N})$, and
\begin{equation}
	(\boldsymbol\alpha_{i},\boldsymbol\alpha_{i}) = (\boldsymbol\alpha_{j},\boldsymbol\alpha_{j}) = -1; \quad \boldsymbol\beta_{ij} = \frac{2(\boldsymbol\alpha_i + \boldsymbol\alpha_j)}{(\boldsymbol\alpha_i + \boldsymbol\alpha_j)^2}.
\end{equation}
The requirement of the perturbation by operators $\mathcal{O}_{ij}$'s to be integrable turns out to be extremely restrictive, almost completely fixing the vectors $\boldsymbol{\alpha}_i$. 

Now we describe the allowed root systems $\{\boldsymbol{\alpha}_i\}$. Consider the Gram matrix $\Gamma_{ij} = (\boldsymbol{\alpha}_i,\boldsymbol{\alpha}_j)$. Since $\Gamma_{ij}$ is symmetric, it can be treated as an incidence matrix of the weighted graph. Since $\Gamma_{ii} = -1$, every node of this graph corresponds to a vector of ''real length'' $\sqrt{-1}$, which we, inspired by the bosonization formula $\psi \leftrightarrow e^{\sqrt{-1}\phi}$, call a \textit{fermionic root} and denote by $\boldsymbol\otimes$. If two fermionic roots $\boldsymbol{\alpha}_i$ and $\boldsymbol{\alpha}_j$ have a non-zero pairing $(\boldsymbol{\alpha}_i, \boldsymbol{\alpha}_j) = x \neq 0$, we draw a line connecting the corresponding nodes and assign a label $x$ to it:
\begin{center}
    \begin{picture}(250,10)(240,123)
        \Thicklines
        \unitlength 5pt
        \put(69,25){\circle{2}}
        \put(79,25){\circle{2}}
        \put(68.4,24,4){\line(1,1){1.2}}
        \put(68.4,25,6){\line(1,-1){1.2}}
        \put(78.4,24,4){\line(1,1){1.2}}
        \put(78.4,25,6){\line(1,-1){1.2}}
        \put(70,25){\line(1,0){8}}
        \put(73.5,26){$x$}
    \end{picture}
\end{center}

The integrability condition requires $\Gamma_{ij}$ to be of a very special form. Specifically, we have the diagrams
\begin{equation}\label{O(2N)-bantik}
\begin{picture}(300,60)(260,110)
    \Thicklines
    \unitlength 5pt
    \put(48,32){\circle{2}}
    \put(48,18){\circle{2}}
    \put(54.4,24,4){\line(-1,-1){7}}
    \put(54.4,25,6){\line(-1,1){7}}
    \put(47.4,31.4){\line(1,1){1.2}}
    \put(47.4,18.6){\line(1,-1){1.2}}
    \put(48,19){\line(0,1){12}}
    \put(55,25){\circle{2}}
    \put(54.4,24,4){\line(1,1){1.2}}
    \put(54.4,25,6){\line(1,-1){1.2}}
    \put(66,25){\line(1,0){8}}
    \put(56,25){\line(1,0){8}}
    \put(65,25){\circle{2}}
    \put(64.4,24,4){\line(1,1){1.2}}
    \put(64.4,25,6){\line(1,-1){1.2}}
    \put(75,25){\circle{2}}
    \put(74.4,24,4){\line(1,1){1.2}}
    \put(74.4,25,6){\line(1,-1){1.2}}
    \put(76,25){\line(1,0){2}}
    \put(80,25){\circle{.2}}
    \put(81,25){\circle{.2}}
    \put(82,25){\circle{.2}}
    \put(83,25){\circle{.2}}
    \put(84,25){\circle{.2}}
    \put(85,25){\circle{.2}}
    \put(87,25){\line(1,0){2}}
    \put(90,25){\circle{2}}
    \put(89.4,24,4){\line(1,1){1.2}}
    \put(89.4,25,6){\line(1,-1){1.2}}
    \put(91,25){\line(1,0){8}}
    \put(100,25){\circle{2}}
    \put(99.4,24,4){\line(1,1){1.2}}
    \put(99.4,25,6){\line(1,-1){1.2}}
    \put(101,25){\line(1,0){8}}
    \put(110,25){\circle{2}}
    \put(109.4,24,4){\line(1,1){1.2}}
    \put(109.4,25,6){\line(1,-1){1.2}}
    \put(109.4,24,4){\line(1,1){8.2}}
    \put(109.4,25,6){\line(1,-1){8.2}}
    \put(117,32){\circle{2}}
    \put(117,18){\circle{2}}
    \put(116.4,32.6){\line(1,-1){1.2}}
    \put(116.4,17.4){\line(1,1){1.2}}
    \put(117,19){\line(0,1){12}}
    \put(55,23){$\underbrace{\phantom{aaaaaaaaaaaaaaaaaaaaaaaaaaaaaaaaaaaaaaaaaaaaa}}$}
    \put(78.5,18){$\scriptstyle{2N-4}$}
    \put(50.5,19){$\scriptstyle{-b^{2}}$}
    \put(50.5,30){$\scriptstyle{-b^{2}}$}
    \put(42,24.5){$\scriptstyle{1+2b^{2}}$}
    \put(57.5,26){$\scriptstyle{1+b^{2}}$}
    \put(68,26){$\scriptstyle{-b^{2}}$}
    \put(103,26){$\scriptstyle{1+b^{2}}$}
    \put(92.5,26){$\scriptstyle{-b^{2}}$}
    \put(111.5,19){$\scriptstyle{-b^{2}}$}
    \put(111.5,30){$\scriptstyle{-b^{2}}$}
    \put(117.5,24.5){$\scriptstyle{1+2b^{2}}$}
    \put(45.5,16){$\scriptstyle{\boldsymbol{\alpha}_{1}}$}
    \put(45.5,33.5){$\scriptstyle{\boldsymbol{\alpha}_{2}}$}
     \put(54.6,27){$\scriptstyle{\boldsymbol{\alpha}_{3}}$}
     \put(64.6,27){$\scriptstyle{\boldsymbol{\alpha}_{4}}$}
     \put(74.6,27){$\scriptstyle{\boldsymbol{\alpha}_{5}}$}
    \put(118,16){$\scriptstyle{\boldsymbol{\alpha}_{2N+1}}$}
    \put(118,33.5){$\scriptstyle{\boldsymbol{\alpha}_{2N+2}}$}
  \end{picture}
  \vspace*{1cm}\\
\end{equation}
for $\mathrm{O}(2N)$ system and
\begin{equation}\label{O(2N+1)-bantik}
\begin{picture}(300,60)(260,110)
    \Thicklines
    \unitlength 5pt
    \put(48,32){\circle{2}}
    \put(48,18){\circle{2}}
    \put(54.4,24,4){\line(-1,-1){7}}
    \put(54.4,25,6){\line(-1,1){7}}
    \put(47.4,31.4){\line(1,1){1.2}}
    \put(47.4,18.6){\line(1,-1){1.2}}
    \put(48,19){\line(0,1){12}}
    \put(55,25){\circle{2}}
    \put(54.4,24,4){\line(1,1){1.2}}
    \put(54.4,25,6){\line(1,-1){1.2}}
    \put(66,25){\line(1,0){8}}
    \put(56,25){\line(1,0){8}}
    \put(65,25){\circle{2}}
    \put(64.4,24,4){\line(1,1){1.2}}
    \put(64.4,25,6){\line(1,-1){1.2}}
    \put(75,25){\circle{2}}
    \put(74.4,24,4){\line(1,1){1.2}}
    \put(74.4,25,6){\line(1,-1){1.2}}
    \put(76,25){\line(1,0){2}}
    \put(80,25){\circle{.2}}
    \put(81,25){\circle{.2}}
    \put(82,25){\circle{.2}}
    \put(83,25){\circle{.2}}
    \put(84,25){\circle{.2}}
    \put(85,25){\circle{.2}}
    \put(87,25){\line(1,0){2}}
    \put(90,25){\circle{2}}
    \put(89.4,24,4){\line(1,1){1.2}}
    \put(89.4,25,6){\line(1,-1){1.2}}
    \put(91,25){\line(1,0){8}}
    \put(100,25){\circle{2}}
    \put(99.4,24,4){\line(1,1){1.2}}
    \put(99.4,25,6){\line(1,-1){1.2}}
    \put(101,25){\line(1,0){8}}
    \put(110,25){\circle{2}}
    \put(109.4,24,4){\line(1,1){1.2}}
    \put(109.4,25,6){\line(1,-1){1.2}}
    \put(109.4,24,4){\line(1,1){8.2}}
    \put(109.4,25,6){\line(1,-1){8.2}}
    \put(117,32){\circle{2}}
    \put(117,18){\circle{2}}
    \put(116.4,32.6){\line(1,-1){1.2}}
    \put(116.4,17.4){\line(1,1){1.2}}
    \put(117,19){\line(0,1){12}}
    \put(55,23){$\underbrace{\phantom{aaaaaaaaaaaaaaaaaaaaaaaaaaaaaaaaaaaaaaaaaaaaa}}$}
    \put(80.5,18){$\scriptstyle{2N-3}$}
    \put(50.5,19){$\scriptstyle{-b^{2}}$}
    \put(50.5,30){$\scriptstyle{-b^{2}}$}
    \put(42,24.5){$\scriptstyle{1+2b^{2}}$}
    \put(57.5,26){$\scriptstyle{1+b^{2}}$}
    \put(68,26){$\scriptstyle{-b^{2}}$}
    \put(103,26){$\scriptstyle{-b^{2}}$}
    \put(92.5,26){$\scriptstyle{1+b^{2}}$}
    \put(110,19){$\scriptstyle{1+b^{2}}$}
    \put(110,30){$\scriptstyle{1+b^{2}}$}
    \put(117.5,24.5){$\scriptstyle{-1-2b^{2}}$}
    \put(45.5,16){$\scriptstyle{\boldsymbol{\alpha}_{1}}$}
    \put(45.5,33.5){$\scriptstyle{\boldsymbol{\alpha}_{2}}$}
     \put(54.6,27){$\scriptstyle{\boldsymbol{\alpha}_{3}}$}
     \put(64.6,27){$\scriptstyle{\boldsymbol{\alpha}_{4}}$}
     \put(74.6,27){$\scriptstyle{\boldsymbol{\alpha}_{5}}$}
    \put(118,16){$\scriptstyle{\boldsymbol{\alpha}_{2n}}$}
    \put(118,33.5){$\scriptstyle{\boldsymbol{\alpha}_{2n+1}}$}
  \end{picture}
  \vspace*{1cm}\\
\end{equation}
for the $\mathrm{O}(2N+1)$ one. Here $b$ is a non-zero Liouville-like parameter.

Given a system of fermionic screening operators, one associates a perturbing operator \eqref{dressed-screening} for each pair $\boldsymbol{\alpha}_i$ and $\boldsymbol{\alpha}_j$ such that $(\boldsymbol{\alpha}_i\cdot\boldsymbol{\alpha}_j)\neq0$. We note that in both cases, $\mathrm{O}(2N)$ and $\mathrm{O}(2N+1)$, there are two ways to define the maximal set of perturbing operators $\mathcal{O}_{ij}$:
\begin{equation}
\begin{aligned}
	\mathbb{I}\,&: \qquad \mathcal{O}_{12}\quad \mathcal{O}_{34}\quad \mathcal{O}_{56}\quad \mathcal{O}_{78}\quad \dots \\
	\mathbb{II}&: \qquad \mathcal{O}_{13}\quad \mathcal{O}_{23}\quad \mathcal{O}_{45}\quad \mathcal{O}_{67} \quad \dots
\end{aligned}
\end{equation}
It can be easily checked that for real $b$'s the exponents $\boldsymbol\beta_{ij}$ 
in the set $\mathbb{I}$ are real and proportional to $b^{-1}$, while the exponents in $\mathbb{II}$ are purely imaginary and proportional to $(i\beta)^{-1}$ with $\beta = \sqrt{1 + b^2}$. It follows that only the operators of the set $\mathbb{I}$ have negative anomalous dimensions and are therefore IR-relevant. In contrast, purely imaginary operators are IR-irrelevant. 

Now, in Toda QFT we expect duality with respect to involution $b^2 \longleftrightarrow -1-b^2$ (or more conveniently $b\longleftrightarrow i\beta$) \cite{Litvinov:2018bou}. Fermionic root systems for $\mathrm{O}(2N+1)$ are clearly self-dual with respect to this involution, while $\mathrm{O}(2N)$ are not. This nontrivial $\mathbb{Z}_2$-action on $\mathrm{O}(2N)$ fermionic root systems is precisely what allows us to define a dual regime in $\mathrm{O}(2N)$ models. Involution $b\longleftrightarrow i\beta$ interchanges real and purely imaginary exponents, and, in this way, it interchanges the sets $\mathbb{I}\longleftrightarrow\mathbb{II}$ comprising the data $\boldsymbol{\beta}_{ij}$ (inner products of relevant exponents). These vectors $\boldsymbol{\beta}_{ij}$'s of $\mathbb{I}$ and $\mathbb{II}$ form affine Lie algebra root systems, which enjoy the same properties with respect to the $b^2 \longleftrightarrow -1-b^2$ involution: $\mathrm{O}(2N+1)$ ones are self-dual while $\mathrm{O}(2N)$ are not. This involution maps $\hat D_{N}^{(2)} \longleftrightarrow \hat D_{N}^{(1)}$ (for $\mathrm{O}(2N)$ models) and $\hat B_N$ to itself (for $\mathrm{O}(2N+1)$):
\begin{equation*}
\psfrag{a}{$\mathrm{O}(2N)$ models:}
\psfrag{b}{usual regime: $\hat D^{(2)}_{N}$ (twisted affine)}
\psfrag{c}{dual regime: $\hat D^{(1)}_N$ (non-twisted affine)}
\psfrag{d}{$\mathrm{O}(2N+1)$ models:}
\psfrag{e}{usual regime: $\hat B_N$}
\psfrag{f}{dual regime: $\hat B_N$}
\includegraphics[width=0.8\textwidth]{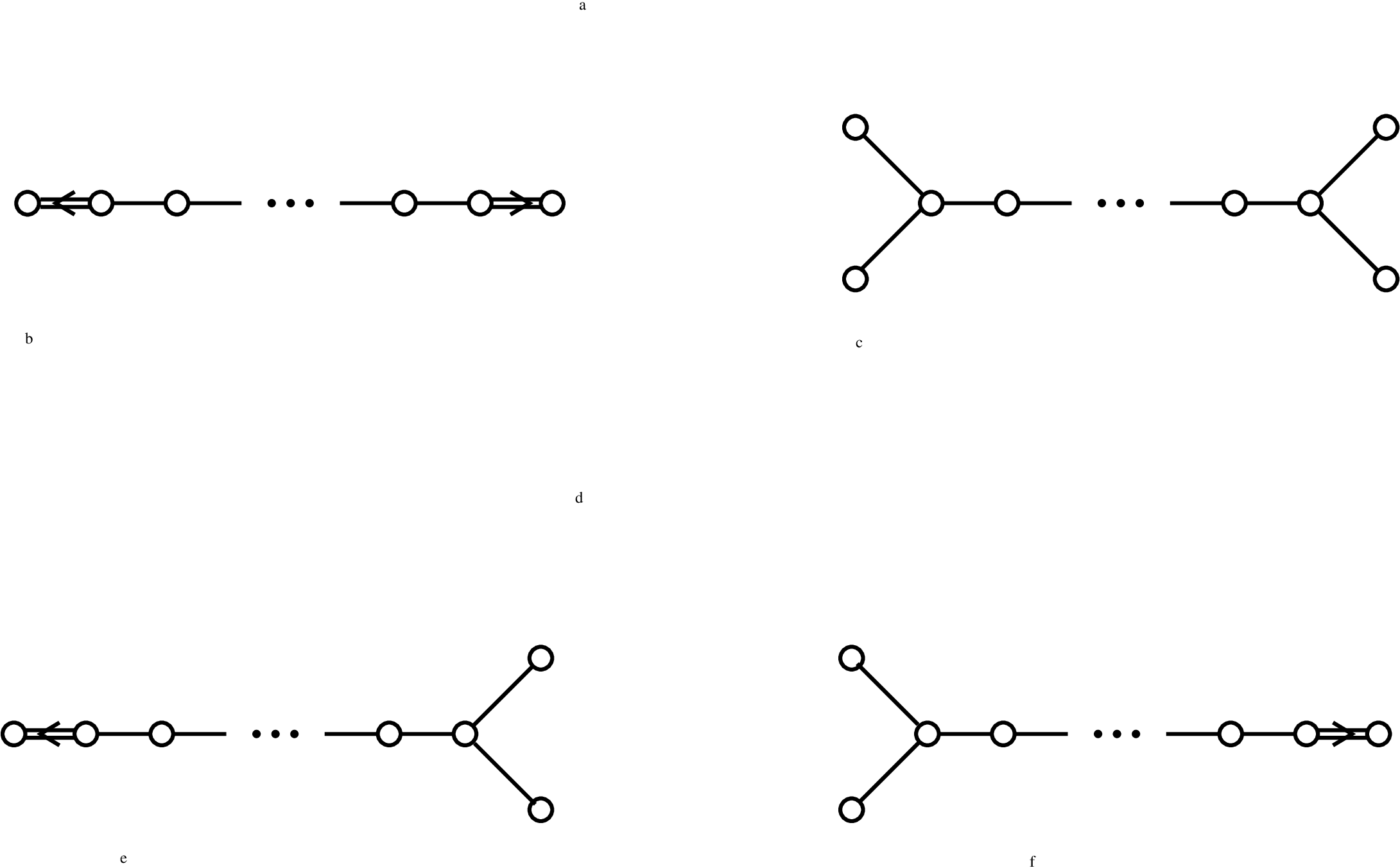}    
\end{equation*}
In both cases the subscript $N$  counts the number of nodes in the diagram minus one (the affine one).

Now, Toda action can be put into a sigma model form
\begin{equation}
    \mathcal{A}_\text{Toda} = \int G_{\mu\nu}(\varphi) \partial\varphi^\mu\overline\partial\varphi^\nu d^2x, \qquad 
    G_{\mu\nu} = \delta_{\mu\nu} + \Lambda \sum\limits_{(i, j)\,\in\,\mathbb{I}\text{\,or\,}\mathbb{II}} (\boldsymbol{\alpha}_i,\partial \boldsymbol{\varphi})(\boldsymbol{\alpha}_i,\overline\partial\boldsymbol{\varphi}) e^{(\boldsymbol{\beta}_{ij},\boldsymbol{\varphi})}.
\end{equation}
A QFT with this action is not renormalizable in a strict sense, since it requires an increasing number of counter-terms at every order of perturbation theory in $\Lambda$. Miraculously, for integrable Toda theories the condition of such a theory to be UV-finite in all loops is presicely the celebrated Ricci flow equation
\begin{equation}
    -\Lambda \frac{d G_{\mu\nu}}{d\Lambda} = R_{\mu\nu} + 2\nabla_\mu\nabla_\nu\Phi.
\end{equation}
Here $\Phi$ is a dilaton field depending on the coordinates of the target space and $\Lambda$, which is found order by order in $\Lambda$. Solving this Ricci flow equation, one arrives at the action of a strongly coupled field theory: a sigma model. Thus, every bowtie diagram gives rise to two renormalizable sigma models, which arise from perturbation by $\mathbb{I}$ and $\mathbb{II}$ correspondingly. 
\section{Main result: dual solution from conformal perturbations}\label{Main}
In the previous section, we explained the relation between certain sigma models and certain Toda field theories. The neighborhood of a UV fixed point of a Yang-Baxter deformation (at least for sphere sigma models) corresponds to a Toda type theory, and vice versa: integrable Toda field theory gives rise to a renormalizable sigma model by imposing Ricci flow. This is undoubtedly a manifestation of a far more general effect, which holds not only for YB-deformed sphere sigma models, but arguably for any integrably deformed sigma model (see \cite{Litvinov:2019rlv,Alfimov:2023evq} for $\mathbb{CP}^n$ models, \cite{Alfimov:2020jpy} for $\mathrm{OSp}(N|2m)$ models). 

In this section, we elaborate on this correspondence and find a series of new solutions to Ricci flow equations, which we call dual regimes of $\mathrm{O}(2N)$ models. Our process is as follows:
\begin{itemize}
	\item Start from a free theory of $2N$ bosonic fields;
	\item Draw a bowtie diagram with $2N$ nodes, choose a regime and add a corresponding integrable perturbation;
	\item Solve a 1-loop RG equation (\ref{1-loop-equations}) for metric assuming no $B$-field is produced. In this step one also has to find the dilaton field $\Phi$;
	\item Perform T-dualities as needed (they do not corrupt Ricci flow). Typically a non-zero pure imaginary $B$-field is produced;
	\item Interpret the resulting metric and $B$-field as those of an integrably deformed sigma model. This deformation can turn out to be the Yang-Baxter one, as in the usual regime, or not.
\end{itemize}
\subsection{Special case: \texorpdfstring{$\mathrm{O}(4)$}{O(4)} model}
Let us implement this general idea on the simplest non-trivial example of $\mathrm{O}(4)$ model.
\subsubsection*{Yang-Baxter regime}
We start from the Toda root system (tetrahedron shape)
\begin{equation}\label{O(4)-bantik}
\begin{picture}(300,60)(140,110)
    \Thicklines
    \unitlength 5pt
    \put(48,32){\circle{2}}
    \put(48,18){\circle{2}}
    \put(62.7,32.7){\line(-1,-1){15.4}}
    \put(49,32.1){\line(1,0){12}}
    \put(49,18){\line(1,0){12}}
    \put(62.8,17.2){\line(-1,1){15.4}}
    \put(47.4,31.4){\line(1,1){1.2}}
    \put(47.4,18.6){\line(1,-1){1.2}}
    \put(48,19){\line(0,1){12}}
    \put(62,32){\circle{2}}
    \put(62,18){\circle{2}}
    \put(61.4,32.6){\line(1,-1){1.2}}
    \put(61.4,17.4){\line(1,1){1.2}}
    \put(62,19){\line(0,1){12}}
    \put(51,20){$\scriptstyle{-b^{2}}$}
    \put(51,29){$\scriptstyle{-b^{2}}$}
    \put(53.5,33){$\scriptstyle{-b^{2}}$}
    \put(53.5,16){$\scriptstyle{-b^{2}}$}
    \put(45.5,16){$\scriptstyle{\boldsymbol{\alpha}_{1}}$}
    \put(45.5,33.5){$\scriptstyle{\boldsymbol{\alpha}_{2}}$}
    \put(62.5,16){$\scriptstyle{\boldsymbol{\alpha}_{3}}$}
    \put(62.5,33.5){$\scriptstyle{\boldsymbol{\alpha}_{4}}$}
    \put(55.5,20){$\scriptstyle{-b^{2}}$}
    \put(55.5,29){$\scriptstyle{-b^{2}}$}
    \put(43,24.5){$\scriptstyle{1+2b^{2}}$}
    \put(62.5,24.5){$\scriptstyle{1+2b^{2}}$}
  \end{picture}
  \vspace*{1cm}
\end{equation}
Denote target-space coordinates by $(x_1, x_2, x_3)$ and write the first order in conformal perturbation theory ($b\rightarrow 0$). In order to do so, it is convenient to introduce an orthonormal basis $\{E_1, e_1, e_2\}$ of $\mathbb{R}^3$ and write for the Toda roots
\begin{equation}
\boldsymbol\alpha_1 = b E_1 + i\beta e_1,
\quad \boldsymbol\alpha_2 = b E_1 - i\beta e_1, 
\quad \boldsymbol\alpha_3 = -b E_1 + i\beta e_2,
\quad \boldsymbol\alpha_4 = -b E_1 - i\beta e_2.
\end{equation}
Corresponding relevant perturbations are $(\boldsymbol\alpha_1,\partial_+x)(\boldsymbol\alpha_1,\partial_-x)e^{(\boldsymbol\beta_{12},\partial_-x)}$ and $(\boldsymbol\alpha_4, \partial_+x)(\boldsymbol\alpha_4,\partial_-x)e^{(\boldsymbol\beta_{34},x)}$, where
\begin{equation}
		\boldsymbol\beta_{12} = \frac{2(\boldsymbol\alpha_1 + \boldsymbol\alpha_2)}{(\boldsymbol\alpha_1+\boldsymbol\alpha_2)^2} = b^{-1}E_1,
        \quad
		\boldsymbol\beta_{34} = \frac{2(\boldsymbol\alpha_3 + \boldsymbol\alpha_4)}{(\boldsymbol\alpha_3+\boldsymbol\alpha_4)^2} = -b^{-1}E_1,
\end{equation}
so, the first order in conformal perturbation theory ($b\rightarrow 0$) for metric and dilaton is
\begin{equation}
	G_{\mu\nu} = \delta_{\mu\nu} + e^{\alpha t}\Big(A_{\mu\nu}\, e^{x_1} + B_{\mu\nu}\,e^{-x_1}\Big) + \dots,\quad
	\Phi = (\rho, x) + \dots,
\end{equation}
where the matrices $A_{\mu\nu}$ and $B_{\mu\nu}$ are
\begin{equation}
	A_{\mu\nu} = (1, i, 0)^{\otimes 2} = 
	\begin{pmatrix}
		1&i&0\\
		i&-1&0\\
		0&0&0
	\end{pmatrix},
	\qquad
	B_{\mu\nu} = (1, 0, i)^{\otimes 2} = 
	\begin{pmatrix}
		1&0&i\\
		0&0&0\\
		i&0&-1
	\end{pmatrix}.
\end{equation}
We note that the metric depends only on the coordinate $x_1$, while $x_2$ and $x_3$ are isometric directions.

Assuming 1-loop RG does not produce nontrivial $B$-field and solving the 1-loop equation
\begin{equation}
	-\dot G_{\mu\nu} = R_{\mu\nu} + 2\,\nabla_\mu\nabla_\nu \Phi
\end{equation}
by iterations, in the first order in $e^{\alpha t}$ we find $\alpha$ and $\rho$ to be
\begin{equation}
	\alpha = \frac{1}{2},\qquad \rho = \left(0;\,-\frac{i}{2};\,\frac{i}{2}\right).
\end{equation}
Higher order terms can be computed in a standard way.  From \cite{Litvinov:2018bou} we know that after T-duality, which produces a purely imaginary $B-$field, and some change of coordinates, we arrive at
\begin{equation}
    ds^2 = \kappa \left[\frac{d\theta_1^2 + \cos^2\theta_1 d\phi_1^2}{1-\kappa^2\sin^2\theta_1} + \sin^2\theta_1 d\chi^2\right],\qquad B = -\frac{i\kappa^2}{2}\left[\frac{\sin2\theta_1}{1-\kappa^2\sin^2\theta_1}d\theta_1\wedge d\phi_1\right]
\end{equation}
which exactly coincides with usual YB solution on $\mathbb{S}^3$. This justifies the duality between the YB-deformed $\mathrm{O}(4)$ model (strong coupling region) and integrably perturbed free CFT (perturbative region).
\subsubsection*{Dual regime}
In the dual regime we start from the dual Dynkin graph ($b^2\leftrightarrow-1-b^2$)
\begin{equation}\label{O(4)-bantik-dual}
\begin{picture}(300,60)(140,110)
    \Thicklines
    \unitlength 5pt
    \put(48,32){\circle{2}}
    \put(48,18){\circle{2}}
    \put(62.7,32.7){\line(-1,-1){15.4}}
    \put(49,32.1){\line(1,0){12}}
    \put(49,18){\line(1,0){12}}
    \put(62.8,17.2){\line(-1,1){15.4}}
    \put(47.4,31.4){\line(1,1){1.2}}
    \put(47.4,18.6){\line(1,-1){1.2}}
    \put(48,19){\line(0,1){12}}
    \put(62,32){\circle{2}}
    \put(62,18){\circle{2}}
    \put(61.4,32.6){\line(1,-1){1.2}}
    \put(61.4,17.4){\line(1,1){1.2}}
    \put(62,19){\line(0,1){12}}
    \put(50.5,19){$\scriptstyle{1+b^{2}}$}
    \put(50.5,30){$\scriptstyle{1+b^{2}}$}
    \put(53.5,33){$\scriptstyle{1+b^{2}}$}
    \put(53.5,16){$\scriptstyle{1+b^{2}}$}
    \put(45.5,16){$\scriptstyle{\boldsymbol{\alpha}_{1}}$}
    \put(45.5,33.5){$\scriptstyle{\boldsymbol{\alpha}_{2}}$}
    \put(62.5,16){$\scriptstyle{\boldsymbol{\alpha}_{3}}$}
    \put(62.5,33.5){$\scriptstyle{\boldsymbol{\alpha}_{4}}$}
    \put(56,19){$\scriptstyle{1+b^{2}}$}
    \put(56,30){$\scriptstyle{1+b^{2}}$}
    \put(41,24.5){$\scriptstyle{-1-2b^{2}}$}
    \put(62.5,24.5){$\scriptstyle{-1-2b^{2}}$}
  \end{picture}
  \vspace*{1cm}
\end{equation}
We introduce coordinates $(x_1, x_2, x_3)$ and orthonormal basis $\{e_1, E_1, E_2\}$ in $\mathbb{R}^3$ such that
\begin{equation}
\boldsymbol{\alpha}_1=i\beta e_1+bE_1,\quad
\boldsymbol{\alpha}_2=i\beta e_1-bE_1,\quad
\boldsymbol{\alpha}_3=-i\beta e_1-bE_2,\quad
\boldsymbol{\alpha}_4=-i\beta e_1+bE_2.
\end{equation}
Relevant perturbations correspond to
\begin{align}
    \boldsymbol{\beta}_{13} &= \frac{2(\boldsymbol\alpha_1 + \boldsymbol\alpha_3)}{(\boldsymbol\alpha_1+\boldsymbol\alpha_3)^2} = \frac{1}{b}(E_1 - E_2) &
    \boldsymbol{\beta}_{14} &= \frac{2(\boldsymbol\alpha_1 + \boldsymbol\alpha_4)}{(\boldsymbol\alpha_1+\boldsymbol\alpha_4)^2} = \frac{1}{b}(E_1 + E_2)\\
    \boldsymbol{\beta}_{23} &= \frac{2(\boldsymbol\alpha_2 + \boldsymbol\alpha_3)}{(\boldsymbol\alpha_2+\boldsymbol\alpha_3)^2} = \frac{1}{b}(-E_1 - E_2) &
    \boldsymbol{\beta}_{24} &= \frac{2(\boldsymbol\alpha_2 + \boldsymbol\alpha_4)}{(\boldsymbol\alpha_2+\boldsymbol\alpha_4)^2} = \frac{1}{b}(-E_1 + E_2)
\end{align}
Thus to the first order the metric and the dilaton read
\begin{equation}
    G_{\mu\nu} = \delta_{\mu\nu} - e^{\alpha t}\Big(A_{\mu\nu}e^{x_1+x_2} + A^\dagger_{\mu\nu} e^{-x_1+x_2} + A_{\mu\nu}e^{x_1-x_2}  + A^\dagger_{\mu\nu} e^{-x_1-x_2}\Big) + \dots\quad
    \Phi = (\rho, x) + ...
\end{equation}
The perturbing matrix is
\begin{equation}
    A_{\mu\nu} = (1, 0, i)^{\otimes 2} =
    \begin{pmatrix}
        1 & 0 & i\\
        0 & 0 & 0\\
        i & 0 & -1
    \end{pmatrix}.
\end{equation}
Assuming that no $B$-field is produced, one solves the 1-loop equation $-\dot G_{\mu\nu} = R_{\mu\nu} + 2\,\nabla_\mu\nabla_\nu \Phi$ and finds $\alpha$ and $\rho$:
\begin{equation}
    \alpha = 1, \qquad \rho = \left(0;\ 0;\ \frac{i}{2}\right).
\end{equation}
Further analysis is rather straightforward. Analyzing first few orders of the expansion of $G_{\mu\nu}(t)$ in powers of $e^{t}$ (in practice first $6$ orders), we arrive at the following ansatz:
\begin{equation}\label{Dual-S3-answer}
    G = \begin{pmatrix}
    \coth F-\frac{\cosh x_1}{\sinh F}&0& -\frac{i\sinh x_1}{\sinh F}\\
    0&\phi&0\\
    -\frac{i\sinh x_1}{\sinh F}&0&\coth F+\frac{\cosh x_1}{\sinh F}
    \end{pmatrix}.
\end{equation}
Analyzing first terms of the UV expansion, one can notice that the expansion of $\phi(t)$ and $F(x_2|t)$ coincides with
\begin{equation}\label{phi-F}
   \phi(t)=-\left(\frac{2K(m)}{\pi}\right)^2\coth\frac{t}{2},\quad
   F(x_2|t)=-\frac{t}{2}+i\,\mathrm{am}(iz|m),
\end{equation}
where
\begin{equation}
    m=-\frac{1}{\sinh^2\frac{t}{2}}\quad\text{and}\quad z=\frac{2K(m)x_2}{\pi},
\end{equation}
and $K(m)$ is the complete elliptic integral of the first kind and $\mathrm{am}(x|m)$ is the Jacobi amplitude.

Now we perform T-duality in $x_3$ direction in \eqref{Dual-S3-answer}. The metric and $B$-field become
\begin{equation}
\begin{aligned}
    ds^2 &= \frac{\sinh F}{\cosh F + \cosh x_1}(dx_1^2 + dx_3^2) +\phi\,dx_2^2\\
    B &= \frac{i \sinh x_1}{\cosh F + \cosh x_1} dx_1\wedge dx_3,
\end{aligned}
\end{equation}
Using the following change of coordinates
\begin{equation}
    \cosh\frac{x_1}{2}=\frac{1}{\cos\theta_2},\quad
    y=i\,\mathrm{am}\left(iz|m\right),\quad
    \tanh\left(\frac{t}{4}-\frac{y}{2}\right)=\tanh\frac{t}{2}\,\sin^2\theta_1,
\end{equation}
the metric and $B$-field in these new coordinates take the following form:
\begin{equation}
\begin{aligned}
    ds^2 &= 4\kappa\left[\frac{d\theta_1^2}{1-\kappa^2\sin^2\theta_1} + \sin^2\theta_1 \frac{d\theta_2^2 + \cos^2\theta_2\,d\chi^2}{1-\kappa^2\sin^4\theta_1\sin^2\theta_2}\right]\\
    B &=-2i\kappa^2\frac{\sin^4\theta_1\,\sin 2\theta_2}{1-\kappa^2\sin^4\theta_1\sin^2\theta_2}d\theta_2\wedge d\chi+4i\tan\theta_2\,d\theta_2\wedge d\chi,
\end{aligned}
\end{equation}
with $ \kappa=-\tanh\frac{t}{2}$.

Let us briefly summarize our findings. There are two ways to integrably perturb the free theory (three-dimensional flat space sigma model): $\mathbb{I}$ and $\mathbb{II}$. The perturbation by $\mathbb{I}$ gives rise to the YB-deformation of $\mathbb S^3$. The metric and $B$-field are
\begin{equation}
\begin{aligned}
    ds^2_{\mathrm{YB}-\mathbb{S}^3} &= \kappa\left[ \frac{d\theta_1^2 + \cos^2\theta_1\,d\phi_1^2}{1 - \kappa^2 \sin^2\theta_1} + \sin^2\theta_1\,d\chi^2 \right]\\
    B_{\mathrm{YB}-\mathbb{S}^3} &= -\frac{i\kappa^2}{2}\frac{\sin2\theta_1}{1-\kappa^2\sin^2\theta_1}d\theta_1\wedge d\phi_1
\end{aligned}
\end{equation}
The second pertubation $\mathbb{II}$, gives rise to a different sigma model on $\mathbb{S}^3$, which is not the YB one. We call this solution dual. Up to a factor of $4$, relabeling $\chi\rightarrow\phi_2$ and dropping the exact form $4i\tan\theta_2\,d\theta_2\wedge d\chi$ the solution is
\begin{equation}
\begin{aligned}
    ds^2_{\mathrm{dual}-\mathbb{S}^3} &= \kappa\left[ \frac{d\theta_1^2}{1 - \kappa^2\sin^2\theta_1} +  \sin^2\theta_1\frac{d\theta_2^2 + \cos^2\theta_2\,d\chi^2}{1 - \kappa^2\sin^4\theta_1\sin^2\theta_2}\right],\\
    B_{\mathrm{dual}-\mathbb{S}^3} &= -\frac{i\kappa^2}{2}\frac{\sin^4\theta_1\,\sin 2\theta_2}{1-\kappa^2\sin^4\theta_1\sin^2\theta_2}d\theta_2\wedge d\chi.
\end{aligned}
\end{equation}
We observe that quite astonishingly the data $(G_{\mu\nu}, B_{\mu\nu}, V_{\mu}, \omega_\mu)$ for both deformed theories on $\mathbb{S}^3$ can be obtained from the data of the YB-deformed theory on $\mathbb{S}^5$:
\begin{equation}\label{YB-metric-S5}
\begin{aligned}
    ds^2_{\mathrm{YB}-\mathbb{S}^5} &= \kappa\left[ \frac{d\theta_1^2 + \cos^2\theta_1\, d\phi_1^2}{1 - \kappa^2\sin^2\theta_1} +  \sin^2\theta_1\frac{d\theta_2^2 + \cos^2\theta_2\,d\phi_2^2}{1 - \kappa^2\sin^4\theta_1\sin^2\theta_2} + \sin^2\theta_1\sin^2\theta_2\,d\chi^2\right]\\
    B_{\mathrm{YB}-\mathbb{S}^5} &= -\frac{i\kappa^2}{2}\left[\frac{\sin2\theta_1}{1-\kappa^2\sin^2\theta_1}d\theta_1\wedge d\phi_1 + \frac{\sin^4\theta_1\,\sin 2\theta_2}{1-\kappa^2\sin^4\theta_1\sin^2\theta_2}d\theta_2\wedge d\phi_2\right]
\end{aligned}
\end{equation}
by setting a pair of coordinates to constants:
\begin{equation}
    \begin{tikzcd}[column sep=5cm]\label{S3-S5-S3-dual}
    & \boxed{\mathrm{YB}-\mathbb{S}^5} \arrow[dl, "\substack{\theta_2 = \pi/2,\\ \phi_2 = \mathrm{const}}"'] \arrow[dr, "\substack{\phi_1 = \mathrm{const},\\ \chi = \mathrm{const}}"] & \\
    \boxed{\text{YB}-\mathbb{S}^3}  &  &  \boxed{\text{dual}-\mathbb{S}^3}
\end{tikzcd}
\end{equation}
\subsection{Special case: \texorpdfstring{$\mathrm{O}(6)$}{O(6)} model}	
Now we turn from $\mathrm{O}(4)$ to $\mathrm{O}(6)$ \cite{Litvinov:2018bou}. Inspired by the observation \eqref{S3-S5-S3-dual}, we will now perform the calculations for $ds^2_{\mathrm{YB}-\mathbb{S}^5}$ and $ds^2_{\mathrm{dual}-\mathbb{S}^5}$ and describe the duality in a similar way: both metrics will arise from YB-deformed theory on $\mathbb{S}^7$ by setting different pairs of coordinates to constants.
\subsubsection*{Usual (Yang-Baxter) regime}
We start from the fermionic Toda root system
\begin{equation}\label{Toda-bowtie-1}
\begin{picture}(300,60)(160,110)
    \Thicklines
    \unitlength 5pt
    \put(48,32){\circle{2}}
    \put(48,18){\circle{2}}
    \put(54.4,24,4){\line(-1,-1){7}}
    \put(54.4,25,6){\line(-1,1){7}}
    \put(47.4,31.4){\line(1,1){1.2}}
    \put(47.4,18.6){\line(1,-1){1.2}}
    \put(48,19){\line(0,1){12}}
    \put(55,25){\circle{2}}
    \put(54.4,24,4){\line(1,1){1.2}}
    \put(54.4,25,6){\line(1,-1){1.2}}
    \put(56,25){\line(1,0){8}}
    \put(65,25){\circle{2}}
    \put(64.4,24,4){\line(1,1){1.2}}
    \put(64.4,25,6){\line(1,-1){1.2}}
    \put(64.4,24,4){\line(1,1){8.2}}
    \put(64.4,25,6){\line(1,-1){8.2}}
    \put(72,32){\circle{2}}
    \put(72,18){\circle{2}}
    \put(71.4,32.6){\line(1,-1){1.2}}
    \put(71.4,17.4){\line(1,1){1.2}}
    \put(72,19){\line(0,1){12}}
    \put(50.5,19){$\scriptstyle{-b^{2}}$}
    \put(50.5,30){$\scriptstyle{-b^{2}}$}
    \put(45.5,16){$\scriptstyle{\boldsymbol{\alpha}_{1}}$}
    \put(45.5,33.5){$\scriptstyle{\boldsymbol{\alpha}_{2}}$}
    \put(55.5,23){$\scriptstyle{\boldsymbol{\alpha}_{3}}$}
    \put(62.5,23){$\scriptstyle{\boldsymbol{\alpha}_{4}}$}
    \put(72.5,16){$\scriptstyle{\boldsymbol{\alpha}_{5}}$}
    \put(72.5,33.5){$\scriptstyle{\boldsymbol{\alpha}_{6}}$}
    \put(66.5,19){$\scriptstyle{-b^{2}}$}
    \put(66.5,30){$\scriptstyle{-b^{2}}$}
    \put(43,24.5){$\scriptstyle{1+2b^{2}}$}
    \put(72.5,24.5){$\scriptstyle{1+2b^{2}}$}
    \put(58.5,26){$\scriptstyle{1+b^{2}}$}
  \end{picture}
  \vspace*{1cm}
\end{equation}
Denote the target-space coordinates by $(x_1, ..., x_5)$ and write perturbed metric and dilaton field in the form $(b\rightarrow 0)$
\begin{equation}
	G_{\mu\nu} = \delta_{\mu\nu} + e^{\alpha t}\Big(A_{\mu\nu}\, e^{x_1} + B_{\mu\nu}\, e^{-x_2} + C_{\mu\nu}\, e^{-x_1 + x_2}\Big) + \dots\quad
	\Phi = (\rho, x) + \dots.
\end{equation}
The matrices $A_{\mu\nu}$, $B_{\mu\nu}$ and $C_{\mu\nu}$ are
\begin{equation}
	A_{\mu\nu} = \big(1, 0, i, 0, 0\big)^{\otimes 2}, \qquad
	B_{\mu\nu} = \big(0, 1, 0, 0, i\big)^{\otimes 2}, \qquad
	C_{\mu\nu} = \big(0, 1, 0, -i, 0\big)^{\otimes 2}.
\end{equation}
Notice that $x_3$, $x_4$, $x_5$ are three isometric directions in the target space by construction.

Assuming 1-loop RG does not produce nontrivial $B$-field and solving the 1-loop equation
\begin{equation}
	-\dot G_{\mu\nu} = R_{\mu\nu} + 2\,\nabla_\mu\nabla_\nu \Phi
\end{equation}
by iterations, in the first order in $e^{\alpha t}$ we find $\alpha$ and $\rho$ to be
\begin{equation}
	\alpha = \frac{2}{3}, \qquad \rho = \frac{1}{6}\,\big(-1, 1, -4i, 2i, 4i\big).
\end{equation}
In \cite{Litvinov:2018bou} the 1-loop Ricci flow was solved exactly (to all powers of $e^{2t/3}$). After some coordinate changes and T-duality in isometric directions (producing non-zero pure imaginary $B$-field) the metric and the $B$-field will reduce exactly to $ds^2_{\mathrm{YB}-\mathbb{S}^5}$ and $B_{\mathrm{YB}-\mathbb{S}^5}$ given by \eqref{YB-metric-S5} (see \cite{Litvinov:2018bou} for details).
\subsubsection*{Dual regime}
Dual regime is obtained by replacing $b^2\longmapsto -1-b^2$ in the fermionic root graph \eqref{Toda-bowtie-1}. 
\begin{equation}
\begin{picture}(300,60)(160,110)
    \Thicklines
    \unitlength 5pt
    \put(48,32){\circle{2}}
    \put(48,18){\circle{2}}
    \put(54.4,24,4){\line(-1,-1){7}}
    \put(54.4,25,6){\line(-1,1){7}}
    \put(47.4,31.4){\line(1,1){1.2}}
    \put(47.4,18.6){\line(1,-1){1.2}}
    \put(48,19){\line(0,1){12}}
    \put(55,25){\circle{2}}
    \put(54.4,24,4){\line(1,1){1.2}}
    \put(54.4,25,6){\line(1,-1){1.2}}
    \put(56,25){\line(1,0){8}}
    \put(65,25){\circle{2}}
    \put(64.4,24,4){\line(1,1){1.2}}
    \put(64.4,25,6){\line(1,-1){1.2}}
    \put(64.4,24,4){\line(1,1){8.2}}
    \put(64.4,25,6){\line(1,-1){8.2}}
    \put(72,32){\circle{2}}
    \put(72,18){\circle{2}}
    \put(71.4,32.6){\line(1,-1){1.2}}
    \put(71.4,17.4){\line(1,1){1.2}}
    \put(72,19){\line(0,1){12}}
    \put(50.5,19){$\scriptstyle{1+b^{2}}$}
    \put(50.5,30){$\scriptstyle{1+b^{2}}$}
    \put(45.5,16){$\scriptstyle{\boldsymbol{\alpha}_{1}}$}
    \put(45.5,33.5){$\scriptstyle{\boldsymbol{\alpha}_{2}}$}
    \put(55.5,23){$\scriptstyle{\boldsymbol{\alpha}_{3}}$}
    \put(62.5,23){$\scriptstyle{\boldsymbol{\alpha}_{4}}$}
    \put(72.5,16){$\scriptstyle{\boldsymbol{\alpha}_{5}}$}
    \put(72.5,33.5){$\scriptstyle{\boldsymbol{\alpha}_{6}}$}
    \put(66,19){$\scriptstyle{1+b^{2}}$}
    \put(66,30){$\scriptstyle{1+b^{2}}$}
    \put(41.8,24.5){$\scriptstyle{-1-2b^{2}}$}
    \put(72.5,24.5){$\scriptstyle{-1-2b^{2}}$}
    \put(58.5,26){$\scriptstyle{-b^{2}}$}
  \end{picture}
  \vspace*{1cm}
\end{equation}
This duality impacts the choice of relevant operators. The dual metric is then
\begin{equation}
	G_{\mu\nu} = \delta_{\mu\nu} - e^{\alpha t}\Big(A_{\mu\nu} e^{x_2 + x_1} + A^\dagger_{\mu\nu} e^{x_2 - x_1} + B_{\mu\nu} e^{-x_2 + x_3} + B^\dagger_{\mu\nu} e^{-x_2 - x_3}\Big) + \dots,
\end{equation}
where the matrices $A_{\mu\nu}^{(\dagger)}$ and $B_{\mu\nu}^{(\dagger)}$ are
\begin{equation}
	A_{\mu\nu} = \big(1, 0, 0, i, 0\big)^{\otimes 2}, \qquad B_{\mu\nu} = \big(0, 0, 1, 0, i\big)^{\otimes 2},
\end{equation}
and $\dagger$ means transpose and complex conjugate. Notice that there are now only two isometric directions ($x_4$, $x_5$), not three as in YB regime.

In parallel to the previously reviewed Yang-Baxter regime, we solve
\begin{equation}
	-\dot G_{\mu\nu} = R_{\mu\nu} + 2\nabla_\mu \nabla_\nu \Phi, \qquad \Phi = (\rho, x) + \dots
\end{equation}
order by order in powers of $e^{\alpha t}$ to find $\alpha$ and $\rho$:
\begin{equation}
	\alpha = 1, \qquad \rho = \frac{1}{2}\left(0, 0, 0, -i, -i\right).
\end{equation}
In \cite{Litvinov:2018bou} the exact solution was found. The metric reads
\begin{equation}\label{Dual-S5-answer}
	\begin{pmatrix}
		\coth F-\frac{\cosh x_{1}}{\sinh F}&0&0&-i\frac{\sinh x_{1}}{\sinh F}&0\\
		0&\phi&0&0&0\\
		0&0&\coth \bar{F}-\frac{\cosh x_{3}}{\sinh\bar{F}}&0&-i\frac{\sinh x_{3}}{\sinh\bar{F}}\\
		-i\frac{\sinh x_{1}}{\sinh F}&0&0&\coth F+\frac{\cosh x_{1}}{\sinh F}&0\\
		0&0&-i\frac{\sinh x_{3}}{\sinh\bar{F}}&0&\coth \bar{F}+\frac{\cosh x_{3}}{\sinh\bar{F}}\\
	\end{pmatrix},
\end{equation}
where similar to \eqref{phi-F}
\begin{equation}
\phi(t)=-\left(\frac{2K(m)}{\pi}\right)^{2}\coth t,\quad \bar{F}(x_{2},t)=F(-x_{2},t),\quad
F(x_2|t)=-t+i\,\textrm{am}(iz|m),
\end{equation}
with
\begin{equation}
	z=\frac{2K(m)}{\pi}\,x_{2},\qquad m=-\frac{1}{\sinh^{2}t}.
\end{equation}
After performing T-duality in $x_4$ and $x_5$ directions and certain peculiar change of coordinates analogous to that of $\mathrm{O}(4)$
\begin{equation*}
\begin{gathered}
    (x_1, x_2, x_3, x_4, x_5) \mapsto (\theta_1, \theta_2, \theta_3, \phi_2, \phi_3),
 \end{gathered}   
\end{equation*}
the metric and  the $B$-field (up to exact form) can be re-casted as 
\begin{equation}\label{S5-dual-metric}
    \begin{aligned}
        ds^2_{\mathrm{dual}-\mathbb{S}^5} &= \kappa\Bigg[\frac{d\theta_1^2}{1-\kappa^2\sin^2\theta_1} + \frac{\sin^2\theta_1\Big(d\theta_2^2 + \cos^2\theta_2\,d\phi_2^2\Big)}{1-\kappa^2\sin^4\theta_1\sin^2\theta_2} + 
        \frac{\sin^2\theta_1\sin^2\theta_2\Big(d\theta_3^2 + \cos^2\theta_3\,d\phi_3^2\Big)}{1-\kappa^2\sin^4\theta_1\sin^4\theta_2\sin^2\theta_3}\Bigg]\\
        B_{\mathrm{dual}-\mathbb{S}^5} &=  -\frac{i\kappa^2}{2}\Bigg[\frac{\sin^4\theta_1\,\sin 2\theta_2}{1-\kappa^2\sin^4\theta_1\sin^2\theta_2}d\theta_2\wedge d\phi_2 + \frac{\sin^4\theta_1\sin^4\theta_2\sin2\theta_3}{1-\kappa^2\sin^4\theta_1\sin^4\theta_2\sin^2\theta_3}d\theta_3\wedge d\phi_3\Bigg]
    \end{aligned}
\end{equation}
with
\begin{equation}
    \kappa(t)=-\tanh t.
\end{equation}

We see that the usual ($\mathbb{I}$) and the dual ($\mathbb{II}$) perturbations give rise to different deformations of $\mathbb{S}^5$: ordinary YB-deformation and newly found dual deformation \eqref{S5-dual-metric}. Moreover, both of them ''lie inside'' the YB-deformed $\mathbb{S}^7$ theory. Indeed, it is straightforward to check that the same rules of setting coordinates from YB-deformed $\mathbb{S}^7$ metric and $B-$field to constants hold for $V$ and $\omega$ as well
\begin{equation}
    \begin{tikzcd}[column sep=5cm]\label{S5-S7-S5-dual}
    & \boxed{\mathrm{YB}-\mathbb{S}^7} \arrow[dl, "\substack{\theta_3 = \pi/2,\\ \phi_3 = \mathrm{const}}"'] \arrow[dr, "\substack{\phi_1 = \mathrm{const},\\ \chi = \mathrm{const}}"] & \\
    \boxed{\text{YB}-\mathbb{S}^5}  &  &  \boxed{\text{dual}-\mathbb{S}^5}
\end{tikzcd}
\end{equation}
\subsection{General conjecture: dual solution on \texorpdfstring{$\mathbb{S}^{2N-1}$}{S(2N)}}
Inspired by the examples of dual solutions on $\mathbb{S}^3$ and $\mathbb{S}^5$, we claim that the data $(ds^2, B, V, \omega)$ on $\mathbb{S}^{2N-1}$ resulting from setting $d\phi_1 = 0$ and $d\chi = 0$ in the usual Yang-Baxter data on $\mathbb{S}^{2N+1}$, solve generalized Ricci flow equations \eqref{1-loop-equations}, and thus provide a new one-parametric integrable deformation of $\mathbb{S}^{2N-1}$ sigma model (dual deformation of $\mathrm{O}(2N)$ sigma-models). More explicitly this dual data reads
\begin{equation}\label{S(2N+1)-DUAL-YB-full-solution}
    ds^{2}_{\text{dual}-\mathbb{S}^{2N-1}} = \frac{\kappa\, d\theta_1^2}{1-\kappa^2\sin^2\theta_1} + \sum_{k=2}^{N}S_{k}(\boldsymbol{\theta})\Bigl(d\theta_{k}^{2}+\cos^{2}\theta_{k}d\phi_{k}^{2}\Bigr),\quad
    B_{\text{dual}-\mathbb{S}^{2N-1}} = \sum_{k=2}^{N}b_{k}(\boldsymbol{\theta})d\theta_{k}\wedge d\phi_{k},
\end{equation}
with
\begin{equation}
\begin{aligned}
    &V_{\text{dual}-\mathbb{S}^{2N-1}} = V_{\text{YB}-\mathbb{S}^{2N}}= -2 \sum\limits_{k=1}^{N} \cot(\theta_{N-k+1})\Big((k-1)F_1(\boldsymbol{\theta}) - \sum\limits_{j=2}^{k} F_j(\boldsymbol{\theta})\Big)d\theta_k\\
    &\omega_{\text{dual}-\mathbb{S}^{2N-1}} = \omega_{\text{YB}-\mathbb{S}^{2N}} = -i\kappa\sum\limits_{k=1}^{n}\big(2N-2k+1\big)\Big(\prod\limits_{j=1}^{k-1}\sin^2\theta_j\Big)\cos^2(\theta_k)\,F_k(\boldsymbol{\theta})\,d\phi_k,
\end{aligned}
\end{equation}
with the same functions $S_{k}(\boldsymbol{\theta})$, $b_k(\boldsymbol{\theta})$, and $F_k(\boldsymbol{\theta})$ as in \eqref{S,S0-ans}, \eqref{b-ans} and \eqref{Fk-ans}. The parameter $\kappa=\kappa(t)$ is the only running coupling
\begin{equation}
    \kappa_{\text{dual}-\mathbb{S}^{2N-1}}(t) = \kappa_{\text{YB}-\mathbb{S}^{2N-1}}(t) = -\tanh(2N-2)t
\end{equation}
This solution is obtained from the YB solution for $\mathbb{S}^{2N+1}$ \eqref{S(2N+1)-YB-full-metric} by setting $\phi_1=\text{const}$ and $\chi=\text{const}$.
\section{Concluding remarks}\label{concl}
In this work, we have defined and studied the dual regime in YB-deformed $\mathrm{O}(2N)$ sigma models. In particular, we have obtained a new series of solutions to the generalized Ricci flow equations \eqref{S(2N+1)-DUAL-YB-full-solution}. We emphasize several important questions and future directions of study.
\begin{itemize}
    \item It is rather natural to expect that dual YB-deformed $\mathrm{O}(2N)$ sigma models should admit a Lagrangian description similar to the description of conventional YB SSSM \eqref{YB-G/H-kappa-action}. Furthermore, the way in which we came to the definition of dual YB-deformed $\mathrm{O}(2N)$ sigma models suggests that they must be classically integrable and admit a Lax representation. Finding such a description is a very interesting and important task.
    \item As the dual YB-deformed $\mathrm{O}(2N)$ sigma models correspond to integrable deformation of $\mathrm{O}(2N)$ model, they are expected to have the same spectrum of fundamental particles as in undeformed $\mathrm{O}(2N)$  models. The scattering matrices of the latter theories have been found in the seminal paper \cite{Zamolodchikov:1978xm}. They coincide up to a total factor with the rational $R-$matrices for $D_N$ algebra. It is well known that there are two non-equivalent trigonometric deformations of $D_N$ $R-$matrices \cite{Bazhanov:1984gu,Jimbo:1985ua} that correspond to affine Dynkin diagrams $\hat{D}_N^{(1)}$ and $\hat{D}_N^{(2)}$. The first deformation has been identified in \cite{Fateev:2018yos} with the YB-deformed $\mathrm{O}(2N)$ model. The identification goes through the construction of Toda field theory whose exponents coincide with the exponents of the fermionic screenings of the bowtie diagram \eqref{O(2N)-bantik} and perturbative computation of the scattering matrix. The dual YB-deformed $\mathrm{O}(2N)$ sigma models are then naturally associated with the second deformation \cite{Bychkov-Nekrasov}.
    \item In the recent paper \cite{Hamidi:2025sgg} a new class of integrable deformations of the principal chiral model called $\mathbb{Z}_n$-twisted trigonometric sigma models has been introduced. Considering in particular the case of $\mathrm{SO}(2N)$ $\mathbb{Z}_2$-twisted models with the twist corresponding to an outer automorphism of $\mathfrak{so}(2N)$ reveals that one of the global symmetry groups gets broken down to $\mathrm{U}(1)^{N-1}$ in contrast with $\mathrm{U}(1)^{N}$ for the usual YB-deformed PCM. In this way, twisted deformed PCM has one less isometries than the untwisted one. The same phenomenon occurs for the dual regime of deformed $\frac{\mathrm{SO}(2N)}{\mathrm{SO}(2N-1)}$ model, and it is therefore natural to expect the existence of a coset version of $\mathbb{Z}_2$-twisted models, which would coincide with the dual regime. 
\end{itemize}
\section*{Acknowledgments}
We acknowledge discussions with Vladimir Bazhanov, Ben Hoare, Boris Nekrasov, Alexey Rosly, Konstantinos Sfetsos and Konstantinos Siampos.

\bibliographystyle{MyStyle}
\bibliography{references}

\providecommand{\href}[2]{#2}\begingroup\raggedright\begin{thebibliography}{10}

\bibitem{Gell-Mann:1960mvl}
M.~Gell-Mann and M.~Levy, \emph{{The axial vector current in beta decay}}, \href{https://doi.org/10.1007/BF02859738}{\emph{Nuovo Cim.} {\bfseries 16} (1960) 705}.

\bibitem{Friedan:1980jm}
D.~H. Friedan, \emph{{Nonlinear Models in Two + Epsilon Dimensions}}, \href{https://doi.org/10.1016/0003-4916(85)90384-7}{\emph{Annals Phys.} {\bfseries 163} (1985) 318}.

\bibitem{Zakharov:1973pp}
V.~E. Zakharov and A.~V. Mikhailov, \emph{{Relativistically Invariant Two-Dimensional Models in Field Theory Integrable by the Inverse Problem Technique}}, {\emph{Sov. Phys. JETP} {\bfseries 47} (1978) 1017}.

\bibitem{Eichenherr:1979ci}
H.~Eichenherr and M.~Forger, \emph{{On the Dual Symmetry of the Nonlinear Sigma Models}}, \href{https://doi.org/10.1016/0550-3213(79)90276-1}{\emph{Nucl. Phys. B} {\bfseries 155} (1979) 381}.

\bibitem{Cherednik:1981df}
I.~V. Cherednik, \emph{{Relativistically Invariant Quasiclassical Limits of Integrable Two-dimensional Quantum Models}}, \href{https://doi.org/10.1007/BF01086395}{\emph{Theor. Math. Phys.} {\bfseries 47} (1981) 422}.

\bibitem{Fateev:1992tk}
V.~A. Fateev, E.~Onofri and A.~B. Zamolodchikov, \emph{{Integrable deformations of the $O(3)$ sigma model. The sausage model}}, \href{https://doi.org/10.1016/0550-3213(93)90001-6}{\emph{Nucl. Phys. B} {\bfseries 406} (1993) 521}.

\bibitem{Fateev:1995ht}
V.~A. Fateev, \emph{{The Duality between two-dimensional integrable field theories and sigma models}}, \href{https://doi.org/10.1016/0370-2693(95)00883-M}{\emph{Phys. Lett. B} {\bfseries 357} (1995) 397}.

\bibitem{Fateev:1996ea}
V.~A. Fateev, \emph{{The sigma model (dual) representation for a two-parameter family of integrable quantum field theories}}, \href{https://doi.org/10.1016/0550-3213(96)00256-8}{\emph{Nucl. Phys. B} {\bfseries 473} (1996) 509}.

\bibitem{Lukyanov:2012zt}
S.~L. Lukyanov, \emph{{The integrable harmonic map problem versus Ricci flow}}, \href{https://doi.org/10.1016/j.nuclphysb.2012.08.002}{\emph{Nucl. Phys. B} {\bfseries 865} (2012) 308} [\href{https://arxiv.org/abs/1205.3201}{{\ttfamily 1205.3201}}].

\bibitem{Klimcik:2008eq}
C.~Klimcik, \emph{{On integrability of the Yang-Baxter sigma-model}}, \href{https://doi.org/10.1063/1.3116242}{\emph{J. Math. Phys.} {\bfseries 50} (2009) 043508} [\href{https://arxiv.org/abs/0802.3518}{{\ttfamily 0802.3518}}].

\bibitem{Delduc:2013fga}
F.~Delduc, M.~Magro and B.~Vicedo, \emph{{On classical $q$-deformations of integrable sigma-models}}, \href{https://doi.org/10.1007/JHEP11(2013)192}{\emph{JHEP} {\bfseries 11} (2013) 192} [\href{https://arxiv.org/abs/1308.3581}{{\ttfamily 1308.3581}}].

\bibitem{Klimcik:2014bta}
C.~Klimcik, \emph{{Integrability of the bi-Yang-Baxter sigma-model}}, \href{https://doi.org/10.1007/s11005-014-0709-y}{\emph{Lett. Math. Phys.} {\bfseries 104} (2014) 1095} [\href{https://arxiv.org/abs/1402.2105}{{\ttfamily 1402.2105}}].

\bibitem{Hoare:2021dix}
B.~Hoare, \emph{{Integrable deformations of sigma models}}, \href{https://doi.org/10.1088/1751-8121/ac4a1e}{\emph{J. Phys. A} {\bfseries 55} (2022) 093001} [\href{https://arxiv.org/abs/2109.14284}{{\ttfamily 2109.14284}}].

\bibitem{Fateev:2018yos}
V.~A. Fateev and A.~V. Litvinov, \emph{{Integrability, Duality and Sigma Models}}, \href{https://doi.org/10.1007/JHEP11(2018)204}{\emph{JHEP} {\bfseries 11} (2018) 204} [\href{https://arxiv.org/abs/1804.03399}{{\ttfamily 1804.03399}}].

\bibitem{Fateev:2019xuq}
V.~Fateev, \emph{{Classical and Quantum Integrable Sigma Models. Ricci Flow, \textquotedblleft{}Nice Duality\textquotedblright{} and Perturbed Rational Conformal Field Theories}}, \href{https://doi.org/10.1134/S1063776119100042}{\emph{J. Exp. Theor. Phys.} {\bfseries 129} (2019) 566} [\href{https://arxiv.org/abs/1902.02811}{{\ttfamily 1902.02811}}].

\bibitem{Litvinov:2018bou}
A.~V. Litvinov and L.~A. Spodyneiko, \emph{{On dual description of the deformed $O(N)$ sigma model}}, \href{https://doi.org/10.1007/JHEP11(2018)139}{\emph{JHEP} {\bfseries 11} (2018) 139} [\href{https://arxiv.org/abs/1804.07084}{{\ttfamily 1804.07084}}].

\bibitem{Alfimov:2020jpy}
M.~Alfimov, B.~Feigin, B.~Hoare and A.~Litvinov, \emph{{Dual description of $\eta$-deformed OSP sigma models}}, \href{https://doi.org/10.1007/JHEP12(2020)040}{\emph{JHEP} {\bfseries 12} (2020) 040} [\href{https://arxiv.org/abs/2010.11927}{{\ttfamily 2010.11927}}].

\bibitem{POLYAKOV1981207}
A.~Polyakov, \emph{Quantum geometry of bosonic strings}, \href{https://doi.org/https://doi.org/10.1016/0370-2693(81)90743-7}{\emph{Physics Letters B} {\bfseries 103} (1981) 207}.

\bibitem{Braaten:1985is}
E.~Braaten, T.~L. Curtright and C.~K. Zachos, \emph{{Torsion and Geometrostasis in Nonlinear Sigma Models}}, \href{https://doi.org/10.1016/0550-3213(86)90196-3}{\emph{Nucl. Phys. B} {\bfseries 260} (1985) 630}.

\bibitem{Squellari:2014jfa}
R.~Squellari, \emph{{Yang-Baxter $\sigma$ model: Quantum aspects}}, \href{https://doi.org/10.1016/j.nuclphysb.2014.02.009}{\emph{Nucl. Phys. B} {\bfseries 881} (2014) 502} [\href{https://arxiv.org/abs/1401.3197}{{\ttfamily 1401.3197}}].

\bibitem{Sfetsos:2015nya}
K.~Sfetsos, K.~Siampos and D.~C. Thompson, \emph{{Generalised integrable \ensuremath{\lambda} - and \ensuremath{\eta}-deformations and their relation}}, \href{https://doi.org/10.1016/j.nuclphysb.2015.08.015}{\emph{Nucl. Phys. B} {\bfseries 899} (2015) 489} [\href{https://arxiv.org/abs/1506.05784}{{\ttfamily 1506.05784}}].

\bibitem{Bykov:2016pfu}
D.~Bykov, \emph{{Complex structure-induced deformations of \ensuremath{\sigma}-models}}, \href{https://doi.org/10.1007/JHEP03(2017)130}{\emph{JHEP} {\bfseries 03} (2017) 130} [\href{https://arxiv.org/abs/1611.07116}{{\ttfamily 1611.07116}}].

\bibitem{Buscher:1987sk}
T.~H. Buscher, \emph{{A Symmetry of the String Background Field Equations}}, \href{https://doi.org/10.1016/0370-2693(87)90769-6}{\emph{Phys. Lett. B} {\bfseries 194} (1987) 59}.

\bibitem{Litvinov:2019rlv}
A.~V. Litvinov, \emph{{Integrable $\mathfrak{gl}(n|n)$ Toda field theory and its sigma-model dual}}, \href{https://doi.org/10.1134/S0021364019230048}{\emph{Pisma Zh. Eksp. Teor. Fiz.} {\bfseries 110} (2019) 723} [\href{https://arxiv.org/abs/1901.04799}{{\ttfamily 1901.04799}}].

\bibitem{Alfimov:2023evq}
M.~Alfimov, I.~Kalinichenko and A.~Litvinov, \emph{{On \ensuremath{\beta}-function of N = 2 supersymmetric integrable sigma-models}}, \href{https://doi.org/10.1007/JHEP05(2024)297}{\emph{JHEP} {\bfseries 05} (2024) 297} [\href{https://arxiv.org/abs/2311.14187}{{\ttfamily 2311.14187}}].

\bibitem{Zamolodchikov:1978xm}
A.~B. Zamolodchikov and A.~B. Zamolodchikov, \emph{{Factorized S matrices in two-dimensions as the exact solutions of certain relativistic quantum field models}}, \href{https://doi.org/10.1016/0003-4916(79)90391-9}{\emph{Annals Phys.} {\bfseries 120} (1979) 253}.

\bibitem{Bazhanov:1984gu}
V.~V. Bazhanov, \emph{{Trigonometric Solution of Triangle Equations and Classical Lie Algebras}}, \href{https://doi.org/10.1016/0370-2693(85)90259-X}{\emph{Phys. Lett. B} {\bfseries 159} (1985) 321}.

\bibitem{Jimbo:1985ua}
M.~Jimbo, \emph{{Quantum R Matrix for the Generalized Toda System}}, \href{https://doi.org/10.1007/BF01221646}{\emph{Commun. Math. Phys.} {\bfseries 102} (1986) 537}.

\bibitem{Bychkov-Nekrasov}
A.~Bychkov and B.~Nekrasov, \emph{to appear},  2025.

\bibitem{Hamidi:2025sgg}
R.~Hamidi and B.~Hoare, \emph{{Twists of trigonometric sigma models}},  \href{https://arxiv.org/abs/2504.18492}{{\ttfamily 2504.18492}}.

\end{thebibliography}\endgroup

\end{document}